 \documentclass[final,3p,times]{elsarticle}

\usepackage{amssymb}
\usepackage{amsthm}
\usepackage{amsmath}

\journal{Elsevier}

\newcommand{\ds}   {\displaystyle}
\newenvironment{enumeri}      
 {\begin{list}{(\roman{enumiii})}
    {\usecounter{enumiii} \setlength{\parsep}{0pt}
     \setlength{\itemsep}{3pt} \settowidth{\labelwidth}{a/}
     \sloppy
    }}{\end{list}}

\begin{document}

\begin{frontmatter}



\title{Microscopic Transport Theory of Nuclear Processes}


\author{K. Dietrich\fnref{permanent}}
\fntext[permanent]
{Permanent address : Physics Department, Technische Universit\"at M\"unchen,
85747 Garching, Germany}
\author{J.-J. Niez}
\author{J.-F. Berger}

\address{CEA, DAM, DIF, F-91297, Arpajon, France}

\begin{abstract}
We formulate a microscopic theory of the decay of a compound nucleus
through fission which generalizes earlier microscopic approaches of fission
dynamics performed in the framework of the adiabatic hypothesis.
It is based on the constrained Hartree-Fock-Bogoliubov procedure and the
Generator Coordinate Method, and requires an effective nucleon-nucleon
interaction as the only input quantity.
The basic assumption is that the
slow evolution of the nuclear shape must be treated explicitely, whereas the
rapidly time-dependent intrinsic excitations can be treated by statistical
approximations. More precisely, we introduce a ``reference density'' $\rho_P$
which represents the slow evolution of the nuclear shape by a reduced density
matrix and the state of intrinsic excitations by a canonical distribution at
each given shape of the nucleus. The shape of the nuclear density
distribution is described by parameters (``generator coordinates'' $q$), not
by ``superabundant'' degrees of freedom introduced in addition to the
complete set of nucleonic degrees of freedom.\\
We first derive a rigorous equation of motion for the reference density
$\rho_P$ and, subsequently, simplify this equation on the basis of the
``Markov approximation''. The temperature $T$ which appears in the canonical
distribution is determined by the requirement that, at each time $t$, the
reference density should correctly reproduce the mean excitation energy at
given values of the shape parameters $q$. The resulting equation for the
``local'' temperature $T(q,t)$ must be solved together with the equations of
motion obtained for the reduced density matrix $R(q_1,q_2;t)$.
\end{abstract}

\begin{keyword}
Nuclear transport theory, fission, HFB, GCM

\PACS \sep 21.60.-n, 21.60.Jz, 24.10.Pa, 24.60.-k, 24.75.+i

\end{keyword}

\end{frontmatter}
\renewcommand{\theequation}{\thesection.\arabic{equation}}

\section{Introduction}     \label{introduction}
\label{Introduction}
\setcounter{equation}{0}
There are many systems in nature which are characterized by more than one
time scale. Let us assume, for simplicity, that there is one time-scale
for the slow, usually collective motion (described by variables or
parameters $q$) and one much shorter time-scale for the rapid motion
of a large number of individual particles (with the degrees of freedom $x$).
Whereas the slow motion must be treated explicitly, the rapid dynamical
evolution can often be approximated by a thermodynamical or, more generally,
by some statistical ``ansatz".
There is a vast literature on the theoretical description of such systems.
One refers to them as ``transport theories".

Their intense study started in the 19$^{th}$ century with the
observation of the ``Brownian motion" \cite{Brown}, i.e. the motion of a pollen
or dust particle submerged in a medium of the randomly moving molecules of a
gas or of a liquid. As an average result of the randomly distributed collisions,
the ``Brownian particles" are exposed to a ``friction force"
$-\gamma\, \overline{\vec{\mathrm v}(t)}$ reducing their average
velocity $\overline{\vec{\mathrm v}(t)}$, and to a dissipative force
measured by a parameter
$D$. The dissipation parameter $D$ is determined by the time rate
$\displaystyle\frac{d}{dt} \overline{\left(\vec{r}(t)-
\overline{\vec{r}(t)}\right)^2}$
of the mean square deviation of the trajectories $\vec{r}(t)$ from the average
trajectory $\overline{\vec{r}(t)}$, the mean values being performed over an
ensemble of trajectories. \\
It was for the model of Brownian motion that Einstein derived the simple
classical relation \cite{Einstein} :
\begin{equation}\label{1.1}
\gamma=\frac{k_B T}{M\,D}
\end{equation}
($M$=mass of the Brownian particle, $T$=temperature, $k_B$=Boltzmann
constant).
The generalized quantum-mechanical version of the Einstein relation (\ref{1.1})
is called ``\underline{fluctuation-dissipation theorem}" \cite{fluc-diss}. It
will be discussed in Section \ref{fluctuation} of our manuscript.

The standard form of transport theories is based on a Hamiltonian $\hat{H}$ of
the form \cite{Hofmann,Grossmann}
\begin{equation}\label{1.2}
\hat{H}=\hat{H}_S(q) + \hat{H}_B(x) +\hat{H}_{SB}(x,q)
\end{equation}
$\hat{H}_S$ depends on the slowly time-dependent variables $q$,
$\hat{H}_B$, often referred to as ``bath Hamiltonian", depends on the rapidly
time-dependent variables $x$, and the Hamiltonian $\hat{H}_{SB}$ represents
the coupling between both kinds of degrees of freedom. \\
In the case of the atomic nucleus and numerous other systems, the
complete set of independent degrees of freedom is given by the rapidly
time-dependent variables of the constituent particles only. Here,
the slow part of the time evolution is produced by a slow collective component
of the particle motion. As an example, a nuclear fission or fusion-fission
process proceeds through the slowly varying shape of the nuclear density
distribution.\\
If, nevertheless, one chooses to describe such a system by a Hamiltonian of type
(\ref{1.2}), the shape variables $q$ are not independent of the
particle variables $x$ but related to them by constraints. Thus, for
establishing a Hamiltonian of type (\ref{1.2}), one needs a theory of
collective motion as a prerequisite. Usually, one adopts a method proposed by
Bohm and Pines \cite{Bohm-Pines}. It is based on the constraint that the
collective variables $q$ are to be equal to mean values of certain multipole
operators $\hat{q}(x)$, which depend on the variables $x$ of the individual
particles :
\begin{equation}\label{1.3}
q= Tr\left(\hat{q}(x)\hat{\rho}\right)
\end{equation}
Here, $\hat{\rho}$ is the density operator of the system or some
approximation of it. Strictly speaking, fluctuations around these mean values
exceed the scope of this approach.

We, henceforth, consider in particular the nuclear fission process. One of
the intriguing aspects of this process is the appearance of
\underline{large} fluctuations \cite{Brosa} of quantum mechanical as well
as thermodynamical origin, as exemplified by the large widths of the mass--
and
kinetic energy--distributions observed even at small excitation energies of
the fissioning system.

In order to remedy this difficulty of the existing dynamical fission theories,
we apply the Generator Coordinate Method (GCM) which has been introduced by Hill
and Wheeler in their well-known paper of 1953 \cite{Hill-Wheeler}.
In this method, the wavefunction $\Psi(x_1, \ldots, x_A ; t)$ of the nucleus is
expanded in terms of a set of basis functions $\phi_a(x_1, \ldots, x_A ; q)$
according to :
\begin{equation} \label{1.4}
\Psi(x_1, \ldots, x_A ; t) = \sum_a \int dq f_a(q,t)
\phi_a(x_1, \ldots, x_A ; q)
\end{equation}
The wave-functions $\phi_a(x_1, \ldots, x_A ; q)$ depend on the nucleon
variables $x_1$, \ldots, $x_A$ and on a limited number of ``generator
coordinates" (g.c.) $q$ which serve to describe the possible shapes
of the nucleus. The subscript ``$a$" defines the quantum states of
intrinsic excitation which may occur for each given shape $q$.
The important aspect of the ansatz (\ref{1.4}) is that the shape parameters $q$
are summed over. Thus, they do not represent ``over-abundant" variables.\\
So far, in actual applications of the GCM to nuclear fission and to nuclear
collective excitations, only the \underline{ground states}
$\phi_{a_0}(x_1, \ldots, x_A ; q)$ of the Hartree-Fock-Bogoliubov (HFB) method
have been superimposed in (\ref{1.4}) \cite{Berger-Gogny,Goutte,Dubray}. These
ground states are defined as the solutions of the variational problem
\begin{equation}\label{1.5}
\delta \langle \phi_{a_0}(q) \vert \hat{H}'\vert \phi_{a_0}(q) \rangle \equiv
\delta \langle \phi_{a_0}(q) \vert
\hat{H} -\lambda \hat{q} -\mu_n \hat{N} -\mu_p \hat{Z}
\vert \phi_{a_0}(q) \rangle =0
\end{equation}
In Eq.(\ref{1.5}), the quantities $\hat{H}$, $\hat{q}$, $\hat{N}$ and $\hat{Z}$
represent the total Hamiltonian, a set of multipole operators, the neutron- and
proton-number operators, respectively. The Lagrange parameters
$\lambda$, $\mu_n$ and $\mu_p$ are defined by the constraints
\begin{align}
&\langle \phi_{a_0}(q) \vert \hat{q}\vert \phi_{a_0}(q) \rangle
 = q    \label{1.6} \\
&\langle \phi_{a_0}(q) \vert \hat{N}\vert \phi_{a_0}(q) \rangle
 = N    \label{1.7} \\
&\langle \phi_{a_0}(q) \vert \hat{Z}\vert \phi_{a_0}(q) \rangle
 = Z    \label{1.8}
\end{align}
where $N$ and $Z$ are the neutron and proton numbers of the nucleus, and $q$ is
a set of generator coordinates.
As usual in the constrained HFB method, the Hamiltonian $\hat{H}$ appearing in
(\ref{1.5}) must be built using an effective nucleon-nucleon interaction
adapted to the description of the intrinsic structure of the mean-field states
$\phi_{a_0}(q)$, such as the Skyrme \cite{Skyrme_force} or the Gogny
\cite{Gogny_force} effective interactions. In Refs.
\cite{Berger-Gogny,Goutte,Dubray}, the Gogny force, which is known to
provide a good description of both the average nuclear mean field and pairing
correlations was used. \\
Omitting intrinsic excitations $\phi_a(q) \neq \phi_{a_0}(q)$ in Eq.(\ref{1.4})
implies that only low-lying vibrational excitations of the system are taken into
account on the way to fission or fusion. Nevertheless, many aspects of the
low-energy fission process like the coexistence of a fusion and a fission
valley \cite{Berger-Gogny} and the basic features of the fragment kinetic
energy distribution \cite{Goutte} have been rather successfully obtained in
spite of this restriction.

In our present work, we want to go beyond these calculations in essentially two
ways :
\begin{enumeri}
\item We include intrinsic excitations, i.e. our Hill-Wheeler basis derived
from constrained HFB calculations includes
intrinsic excitations $\phi_a(q)$ up to a given upper energy limit.
\item We represent the system by the density operator $\hat{\rho}(t)$ instead
of describing it by a single state vector $\Psi(t)$.
\end{enumeri}
As excited states $\phi_a(q)$, we use the eigenstates of the one-body part
$\hat{H}'_{1}(q)$ in the Wick decomposition of the Hamiltonian $\hat{H}'(q)$
(see Eq. (\ref{1.10})).\\
As the part $\hat{H}'^{(20)}+\hat{H}'^{(02)}$ of the Wick decomposition
involving the creation ($\hat{H}'^{(20)}$) or annihilation ($\hat{H}'^{(02)}$)
of two quasi-particles vanishes as a result of the variational principe
(\ref{1.5}), the Hamiltonian $\hat{H}'(q)$ reads :
\begin{equation}\label{1.9}
\hat{H}'(q) =H'^{(0)}(q)+\hat{H}'^{(11)}(q)+\hat{H}'_{(2)}(q)
\end{equation}
The Hamiltonian $\hat{H}'^{(11)}(q)$ comprises all the terms with one creation
and one annihilation operator, and $\hat{H}'_{(2)}(q)$ is defined as the sum
of all terms involving four quasi-particle creation or annihilation operators
in normal order. Hence:
\begin{equation}\label{1.10}
\hat{H}'_1(q) \phi_a(q) \equiv \left(  H'^{(0)}(q)+\hat{H}'^{(11)}(q)\right)
 \phi_a(q)  = E'_{1a}(q) \phi_a(q)
\end{equation}
The ``vacuum energy" at a given $q$ is given by
\begin{equation}\label{1.11}
E'_{1a_0}(q) =  H'_{(0)}(q)
\end{equation}
with the corresponding vacuum state $\phi_{a_0}(q)$ (see Eq.(\ref{1.5})).
A presumably preferable choice of $\phi_{a_0}(q)$  would be provided by the
solutions of the RPA at a given value of the g.c. $q$. Given the
fact that it is difficult to determine these \underline{multiphonon states}
for a heavy
nucleus, we shall use the states defined in Eq.(\ref{1.10}) which represent
\underline{multi-quasiparticle states} in a nucleus with the mean multipole
moments $q$.

Two well-known practical difficulties are inherent to the GCM :
\begin{enumeri}
\item The Hill-Wheeler states $\phi_a(q)$ are not orthogonal to each other.
\item There may occur a number of linearly dependent or almost linearly
dependent states among the basis states $\phi_a(q)$. This is revealed by the
fact that the overlap matrix
\begin{equation}\label{1.12}
N_{a_1 a_2}(q_1,q_2) := \langle \phi_{a_1}(q_1) \vert \phi_{a_2}(q_2) \rangle
\end{equation}
may have a number of vanishing or almost vanishing eigenvalues.
\end{enumeri}
We apply two remedies to cope with these difficulties: \\
1) We work with \underline{discrete values of the generator coordinates}
(g.c.) instead of the continuous range of values implied in Eq.(\ref{1.4}):
Each given type of g.c. $q$ varying from a lower limit
$inf(q)$ to an upper limit $sup(q)$ is replaced by a finite number of
discrete values with a spacing $\Delta q$:
\begin{equation}\label{1.13}
q \rightarrow \inf(q)+n \Delta q \leq \sup(q)
\end{equation}
where the positive integer numbers $n$ range from 0 to $(sup(q)
-inf(q))/\Delta q$. The step length $\Delta q$ must be chosen in a physically
reasonable way
\begin{equation}\label{1.14}
\int dq \rightarrow \Delta q \sum_{n} \equiv \sum_{q}
\end{equation}
As we incorporate intrinsic excitations ``a" for each set of the g.c. $q$, we
surmise that the required accuracy of description can be achieved with not
too small values of the step length $\Delta q$.

2) We either use a biorthogonal system of basis states
$\widetilde{\phi_a(q)}$, $\phi_a(q)$ with
\begin{equation}\label{1.15}
\langle\widetilde{\phi_{a_1}(q_1)} \vert \phi_{a_2}(q_2)\rangle
=\delta_{a_1a_2} \delta_{q_1q_2}
\end{equation}
or we replace the non-orthogonal Hill-Wheeler states $\phi_a(q)$ by a system
of orthonormal basis functions $\psi_a(q)$.

Our paper is organized as follows: \\
The just mentioned alternative of either
using ``adjoint" (=``dual") basis states $\widetilde{\phi_a(q)}$ as
``bra-states" in all the matrix elements or of introducing an adequately
chosen orthonormal set of basis states $\psi_a(q)$ will be dealt with in
\underline{Section \ref{choice}}.\\
In \underline{Section \ref{derivation}}, we shall introduce our basic
approximation $\hat{\rho}_P(t)$ for the statistical operator
$\hat{\rho}(t)$ of the total system and derive a rigorously valid equation
of motion for $\hat{\rho}_P(t)$ using the formalism introduced by Nakajima
and Zwanzig \cite{Nakajima, Zwanzig}.
This rigorously valid equation cannot be solved as it stands. An important
and well-known step in transport theories is the application of the
so-called ``\underline{Markov approximation}". Applying this approximation,
we arrive at an integro-differential equation for $\hat{\rho}_P(t)$  which
is amenable to a numerical solution.\\
The physical content of the Markov approximation is that the slowly
time-dependent part $\hat{\rho}_P(t)$  of the statistical operator
$\hat{\rho}(t)$, which is designed to describe the slow collective part of
the time evolution, should only depend in a negligible way on the
detailed phase relations of the rapidly time-dependent intrinsic variables.
The detailed form of the Markov approximation will be treated in
Section \ref{derivation}.\\
In the same Section, we shall also introduce a decomposition of the total
Hamiltonian $\hat{H}$ into an average collective potential $\hat{H}_{coll}$
and a remainder
$\hat{H}_{cpl}=\hat{H} - \hat{H}_{coll}$, and we shall postulate
that the coupling Hamiltonian $\hat{H}_{cpl}$ can be treated in
perturbation theory.

Within the slowly time-dependent part $\hat{\rho}_P(t)$ of the density
operator $\hat{\rho}(t)$, the distribution of the system over the
intrinsic excitations at given value of the g.c. $q$ will be described by a
canonical distribution which depends on a temperature $T$.
This temperature $T(q,t)$, which should be considered as a useful parameter,
will be chosen as a function of the generator
coordinate $q$ and of the time $t$ in such a way that
the ansatz $\hat{\rho}_P(t)$ represents optimally the slowly time-dependent
part of the density operator $\hat{\rho}(t)$. The choice of
the temperature $T(q,t)$ and the equation by which it is to be determined
will be presented in \underline{Section \ref{determination}}.

It is a well-known feature of the Generator Coordinate Method (GCM) that
matrix-elements of the Hamiltonian or of other observables $\hat{O}$
between Hill-Wheeler states $\phi_{a_1}(q_1)$  and $\phi_{a_2}(q_2)$ contain
a factor which depends in a narrow, approximately Gaussian form
on the difference $r$=$q_1-q_2$, whereas the remaining part of the matrix
element is a smooth function of the g.c. $q_1$ and $q_2$.
Writing the elements of the overlap matrix in the form
\begin{equation}\label{1.16}
N_{a_1a_2}(q_1,q_2) = \langle \phi_{a_1}(q_1)\vert\phi_{a_2}(q_2) \rangle
= e^{-r^2/2\sigma^2} n_{a_1a_2}(q_1,q_2),
\end{equation}
the factor $n_{a_1a_2}(q_1,q_2)$ depends smoothly on the variables
$q_1$ and $q_2$. Analogously, it is useful to define a
``reduced matrix element"
 $\langle \phi_{a_1}(q_1)\vert\hat{o}\vert\phi_{a_2}(q_2) \rangle$
 by the equation
\begin{equation}\label{1.17}
\langle \phi_{a_1}(q_1)\vert\hat{o}\vert\phi_{a_2}(q_2) \rangle
:=\frac{\langle \phi_{a_1}(q_1)\vert\hat{O}\vert\phi_{a_2}(q_2) \rangle}
{N_{a_1 a_2}(q_1,q_2)}
\end{equation}
It is a convenient and usually good approximation \cite{GOA} to
assume that the reduced matrix-elements
$\langle \phi_{a_1}(q_1)\vert\hat{o}\vert\phi_{a_2}(q_2) \rangle$ $\equiv$
$\langle \phi_{a_1}(q_{12}+r/2)\vert\hat{o}\vert$ $\phi_{a_2}(q_{12}-r/2)
\rangle$
be represented by a Taylor expansion around $q_1=q_2=q_{12}\equiv (q_1+q_2)/2$
up to second order.\\
In transport theories, this approximation is customarily called the
``\underline{Fokker-Planck approximation (FPA)}".
It was introduced in order to replace the ``master equation" which is an
integro-differential equation for a classical probability distribution
depending on slowly varying variables, by simple differential equations
of second order, the so-called ``Fokker-Planck equations (FPE)". A pertinent
presentation of this procedure is given in Ref. \cite{vanKampen}.

\underline{Section \ref{fluctuation}} of our paper will be devoted to the
fluctuation-dissipation
relation: It will be shown that, in the framework of our approach, such a
relation holds for those transport coefficients in the NZ equation which are
multiplied with the ``diagonal" part $R(q_1,q_1;t)$ of the reduced density
matrix $R(q_1,q_2;t)$ (see Eq. (\ref{5.10})). Furthermore, as will become
clear, the derivation of the fluctuation-dis\-sip\-ation relation relies on
a few additional approximations beyond the mere perturbative treatment of the
coupling Hamiltonian $\hat{H}_{cpl}$.

As we take into account the occurrence of intrinsic excitations of the
fissioning or fusing system, our theory may be applied up to moderately
high excitation energies.
The upper limit of its applicability will be given by the appearance of final
channels with more than two final composite particles, in particular, the
\underline{``incomplete fusion" channels}. In these reactions, only a part
of the incoming projectile nucleus fuses and, subsequently, undergoes
fission, whereas the remaining part is immediately leaving the interaction zone in
predominantly forward direction. \\
The emission of neutrons, protons, and photons by the fissioning nucleus will
be described within the framework of our theory in a paper which we shall
submit in due time to the same journal.
For excitation energies in excess of about 50 MeV, the evaporation of
neutrons and protons has been rather successfully described on the basis of
the Thomas-Fermi approximation and a simple Fokker-Planck equation for the
fissioning nucleus \cite{KDKP}.\\
Whereas the intrinsic excitations can only be treated quantum-mechanically,
the collective motion leading to fission via the passage through strongly
deformed nuclear shapes can approximately be described as a classical motion,
at least at excitation energies where tunnelling is negligible. Indeed,
valuable insights into the fusion- and fission- processes have been obtained
within classical models \cite{Swiatecki-Blocki}.\\
In our theory, the classical limit could be investigated by introducing the
Wigner-Weyl representation for the statistical operator and by replacing
commutators by Poisson brackets. We refrain from going into these details in
the present paper.

In Section \ref{limits}, we discuss the relation of our theory to earlier work
on the microscopic theory of nuclear fission.

In Section \ref{summary}, we summarize our results and point
out open problems.

\section{Choice of basis functions}     \label{choice}
\setcounter{equation}{0}

The choice of the basis-functions is of great importance in our theory,
because their dependence on the generator coordinates $q$ enters our
description of the slow collective motion and of the repartition of the
intrinsic excitation energy.In the two subsections of this chapter, we deal
with the biorthogonal basis and with a physically motivated special choice
of an orthonormal basis.

The Hill-Wheeler (HW) states introduced in Eq. (\ref{1.10}) span a
${\cal N}$-dim\-ensio\-nal vector space. If ${\cal N}_{gc}$ is the number of
discrete values of the generator coordinates and if one associates a given
number ${\cal N}_{intr}$ of intrinsic excitations with each given value of
the generator coordinates, the dimension of the vector space is given by
${\cal N}={\cal N}_{gc}\cdot{\cal N}_{intr}$. \\
The ${\cal N}$ Hill-Wheeler states $\phi_a(q)$ and their adjoints
$\widetilde{\phi_a(q)}$ defined by the property (\ref{1.15}) satisfy the
completeness relation
\begin{equation}\label{2.1}
\sum_{a,q}  \vert \phi_a(q) \rangle \langle \widetilde{\phi_a(q)} \vert=
\sum_{a,q}  \vert \widetilde{\phi_a(q)} \rangle \langle \phi_a(q) \vert= 1
\end{equation}
Henceforth, we convene to use the ordinary HW-states as ``ket"-states and
the adjoints as ``bra"-states. \\
The adjoint states are obtained from the ordinary HW-states through the
inverse $\left(N^{-1}\right)_{a_1a_2}(q_1,q_2)$ of the overlap matrix
(\ref{1.12}), as can be easily seen :
\begin{align}
&\langle \widetilde{\phi_{a_1}(q_1)} \vert
=\sum_{a_2q_2}\langle \widetilde{\phi_{a_1}(q_1)} \vert
 \widetilde{\phi_{a_2}(q_2)}  \rangle \langle \phi_{a_2}(q_2) \vert
\label{2.2}\\
&\langle \widetilde{\phi_{a_1}(q_1)} \vert  \widetilde{\phi_{a_2}(q_2)}
 \rangle = \left(N^{-1}\right)_{a_1a_2}(q_1,q_2)
\label{2.3}
\end{align}
The labels $a_1,q_1$ characterizing a given adjoint state
$\widetilde{\phi_{a_1}(q_1)}$ denote the state with the largest amplitude
$\left(N^{-1}\right)_{a_1a_2}\!\!(q_1,q_2)$ in Eq. (\ref{2.2}). We write the
overlap matrix in the form
\begin{equation}\label{2.4}
N=1+K
\end{equation}
The eigenvalues of the matrix $K$ lie within the unit circle. Consequently,
the series expansion for the inverse $N^{-1}$ converges
\begin{equation}\label{2.5}
N^{-1} = (1+K)^{-1}= 1-K+K^2-K^3+\ldots
\end{equation}
The order of magnitude of the term $K^n$ in Eq. (\ref{2.5}) is given by
$\varepsilon^n$ where the ``overlap parameter" $\varepsilon$ is defined by
\begin{equation}\label{2.6}
\varepsilon := \exp{(-\Delta q^2/2\sigma_0^2)}
\end{equation}
The parameter $\Delta q$ is the distance between neighbouring discrete values
of the generator coordinate $q$. To each type of generator coordinate
corresponds a certain parameter $\varepsilon$.\\
In Fig.1, we present the overlap matrix $N$ for the simple case of 6
different HW-states.

\begin{figure}[ht!]\label{fig1}
\begin{center}
\begin{tabular}{rrccc}
\hspace*{10mm}&
&$\!\!\!\!\!\!\overbrace{\hspace{18mm}}^{\textstyle q^{(1)}}$
&$\!\!\!\overbrace{\hspace{18mm}}^{\textstyle q^{(2)}}$
&$\!\!\!\overbrace{\hspace{18mm}}^{\textstyle q^{(3)}}$\\
\end{tabular} \\
\begin{tabular}{rr|c|c|c|}
 \cline{3-5}
 $q^{(1)}$ & $\left\{\rule{0mm}{7mm} \right.$ &
 1 & $N(q^{(1)},q^{(2)})$  & $N(q^{(1)},q^{(3)})$ \\
 \cline{3-5}
 $q^{(2)}$ & $\left\{\rule{0mm}{7mm} \right.$ &
   $N(q^{(2)},q^{(1)})$  & 1 & $N(q^{(2)},q^{(3)})$ \\
 \cline{3-5}
 $q^{(3)}$ & $\left\{\rule{0mm}{7mm} \right.$ &
   $N(q^{(3)},q^{(1)})$  & $N(q^{(3)},q^{(2)})$ &  1 \\
 \cline{3-5}
\end{tabular}
\end{center}
\caption{Form of the overlap matrix $N$ for the case of 3 values
($q^{(1)}$, $q^{(2)}$, $q^{(3)}$) of the g.c. and 2 states
of intrinsic motion ($a^{(1)}$, $a^{(2)}$). The $2\times 2$ submatrices for
given values $q^{(i)}$$\neq$$ q^{(j)}$ and $q^{(i)}$=$ q^{(j)}$ of the g.c.
are denoted by $N(q^{(i)},q^{(j)})$ and $1$, respectively.}
\end{figure}
The matrix-elements of a hermitian operator $\hat{O}=\hat{O}^\dagger$ between
biorthogonal states satisfy the condition
\begin{equation}\label{2.7}
O_{a_1a_2}(q_1,q_2) :=
\langle \widetilde{\phi_{a_1}(q_1)}\vert\hat{O} \vert\phi_{a_2}(q_2) \rangle
= \langle \phi_{a_2}(q_2)\vert\hat{O} \vert\widetilde{\phi_{a_1}(q_1)} \rangle^*
\end{equation}
The matrix $O\equiv\left\{O_{a_1a_2}(q_1,q_2)\right\}$ is thus not
hermitian. The lack of hermiticity of the biorthogonal matrix representation
of a hermitian operator $\hat{O}$ is of no consequence for the thermodynamic
mean value
\begin{equation}\label{2.8}
\langle \langle \hat{O} \rangle \rangle :=
{\rm Tr}\left(\hat{\rho}(t)\hat{O}\right)
\end{equation}
given the fact that the trace is independent of the representation in which
it is evaluated.

Let us note that the adjoint basis-states $\widetilde{\phi_a(q)}$ have no
physical meaning and that the operators
\begin{equation}\label{2.9}
\hat{\Pi}_a(q) :=
\vert \phi_a(q)\rangle\langle\widetilde{\phi_a(q)}\vert
\end{equation}
although satisfying the property
\begin{equation}\label{2.10}
\hat{\Pi}_a(q)  \hat{\Pi}_{a'}(q') =\delta_{aa'}\delta_{qq'}\hat{\Pi}_a(q)
\end{equation}
are not projection operators in the usual sense, as they are not hermitian.

One can avoid the inconveniences of the representation in terms of
bior\-tho\-go\-nal basis functions by introducing a complete system of
orthonormal
basis states in the ${\cal N}$-dimensional HW-space. There are, of course,
infinitely many possible choices of orthonormal basis functions. One
possibility would be to use the orthonormal
eigenstates of the overlap matrix $N$ as a system of basis functions.\\
A more physical choice is to determine the eigenstates of the hermitian
operator
\begin{equation}\label{2.11}
\hat{H}^{M} := \sum_{a_1q_1a_2q_2}
\vert  \widetilde{\phi_{a_1}(q_1)} \rangle
{\cal H}^{M}_{a_1a_2}(q_1,q_2)\langle\widetilde{\phi_{a_2}(q_2)} \vert
\end{equation}
where the hermitian matrix ${\cal H}^{M}_{a_1a_2}(q_1,q_2)$ is defined by
\begin{equation}\label{2.12}
{\cal H}^{M}_{a_1a_2}(q_1,q_2) := \frac{1}{2} \left( E'_{1a_1}(q_1)+
E'_{1a_2}(q_2)\right) N_{a_1a_2}(q_1,q_2)
\end{equation}
The energies $E'_{1a}(q)$ pertain to states of independently  moving
quasi-particles in a nucleus whose shape is defined by the g.c. $q$ (see Eq.
(\ref{1.10})). The orthonormal eigenstates $\Psi_\alpha$ of the Hamiltonian
(\ref{2.11}) with eigenvalues $\varepsilon_\alpha$
\begin{equation}\label{2.13}
\langle \Psi_\beta\vert\hat{H}^{M} \vert\Psi_\alpha \rangle
=\varepsilon_\alpha  \langle \Psi_\beta\vert\Psi_\alpha \rangle
=\varepsilon_\alpha  \delta_{\alpha\beta}
\end{equation}
can be expanded in terms of the HW-states $\phi_a(q)$ or their adjoints
\begin{equation}\label{2.14}
\vert\Psi_\alpha \rangle =\sum_{a',q'}\vert\phi_{a'}(q') \rangle
\langle \widetilde{\phi_{a'}(q')}\vert \Psi_\alpha \rangle
=\sum_{a',q'}\vert\widetilde{\phi_{a'}(q')} \rangle
\langle \phi_{a'}(q') \vert \Psi_\alpha \rangle
\end{equation}
In the limit of vanishing overlap parameter $\varepsilon$, a given eigenstate
$\Psi_\alpha$ becomes equal to a specific HW-state $\phi_a(q)$. Thus, we can
denote the orthonormal basis states $\Psi_\alpha$  by $\Psi_a(q)$ where the
labels $a,q$ characterize the main component  $\phi_a(q)$ in Eq. (\ref{2.14})
which survives for $\varepsilon\rightarrow 0$.   \\
The coefficients
\begin{align}
& D^\beta_{aq} :=\langle \phi_a(q) \vert\Psi_\beta \rangle \label{2.15}\\
& \widetilde{D}_\alpha^{aq} :=\langle \widetilde{\phi_a(q)} \vert\Psi_\alpha
\rangle \label{2.16}
\end{align}
satisfy the relations
\begin{align}
& \sum_{aq} D^{\beta^*}_{aq}  \widetilde{D}_\alpha^{aq}=
\langle \Psi_\beta\vert\Psi_\alpha \rangle = \delta_{\alpha\beta}
\label{2.17}\\
& \sum_{\alpha} \widetilde{D}_\alpha^{a_2q_2} D^{\alpha^*}_{a_1q_1}
 =\delta_{a_1a_2} \delta_{q_1q_1} \label{2.18}\\
& D^\beta_{a_1q_1} =\langle \phi_{a_1}(q_1) \vert\Psi_\beta \rangle
=\sum_{a_2q_2} N_{a_1a_2}(q_1,q_2) \widetilde{D}_\beta^{a_2q_2}
   \label{2.19}\\
& \widetilde{D}_\beta^{a_1q_1} =  \langle \widetilde{\phi_{a_1}(q_1)}
\vert\Psi_\beta \rangle
=\sum_{a_2q_2} \left(N^{-1}\right)_{a_1a_2}(q_1,q_2)D^\beta_{a_2q_2}
   \label{2.20}
\end{align}
Eq. (\ref{2.13}) can be written in the form
\begin{equation}\label{2.21}
\sum_{a_1q_1a_2q_2}
\widetilde{D}_\beta^{{a_1q_1}^*} {\cal H}^{M}_{a_1a_2}(q_1,q_2)
\widetilde{D}_\alpha^{a_2q_2} =  \varepsilon_\alpha  \delta_{\alpha\beta}
\end{equation}
and the eigenenergies $\varepsilon_\alpha$ can also be obtained as solutions
of the equation
\begin{equation}\label{2.22}
{\rm det}\{{\cal H}^{M}_{a_1a_2}(q_1,q_2) - \varepsilon_\alpha
N_{a_1a_2}(q_1,q_2)\} =0
\end{equation}
In the case that the overlap parameter $\varepsilon$ is noticeably smaller
than 1, a given eigenstate $\Psi_\alpha = \Psi_a(q)$ of $\hat{H}^{M}$
describes physically a state of independent quasi-particles whose main
component is the eigenstate $\phi_a(q)$ of the HB-Hamiltonian
$\hat{H}'_1(q)$. However, it contains admixtures of eigenstates
$\phi_{a'}(q')$ of the HB-Hamiltonian $\hat{H}'_1(q')$ with
$q'=q\pm \Delta q,q\pm 2\Delta q,\ldots $

Let us note that the operator defined by
\begin{equation}\label{2.23}
\hat{\Pi}_\alpha = \vert \Psi_\alpha\rangle \langle\Psi_\alpha \vert
= \vert  \Psi_a(q)\rangle \langle \Psi_a(q)\vert
\end{equation}
is a true projection operator satisfying not only the relations
\begin{equation}\label{2.24}
\hat{\Pi}_\alpha \hat{\Pi}_\beta =\delta_{\alpha\beta} \hat{\Pi}_\alpha
\end{equation}
but also
\begin{equation}\label{2.25}
\hat{\Pi}_\alpha = \hat{\Pi}_\alpha ^\dagger
\end{equation}

As most of the results to be obtained in the following chapters hold in the
same form for the matrix-representation in terms of the biorthogonal or the
orthonormal basis-functions, we introduce a unified notation encompassing
the two alternatives in the following table
\renewcommand{\arraystretch}{1.5}
\begin{center}
\begin{tabular}{|c|c|c|}
 \hline
unified & biorthogonal & orthonormal \\ [-3mm]notation & basis & basis\\
 \hline
$\vert a q\rangle$ & $\vert\phi_a(q)\rangle$ & $\vert\Psi_a(q)\rangle$\\
 \hline
$\langle a q\vert$ & $\langle \widetilde{\phi_a(q)}\vert$ &
$\langle \Psi_a(q)\vert$\\
 \hline
 ${\cal E}'_a(q)$ &  $E'_{1a}(q)$  & $\varepsilon'_\alpha(q)$\\
 \hline
& &  \\ [-7mm]
$\langle a_1q_1\vert\hat{O} \vert a_2q_2 \rangle$  &
$\langle \widetilde{\phi_{a_1}(q_1)}\vert\hat{O} \vert\phi_{a_2}(q_2)
\rangle$
&  $\langle \Psi_{a_1}(q_1)\vert\hat{O} \vert\Psi_{a_2}(q_2) \rangle$ \\
 \hline
\end{tabular}
\end{center}

\section{Derivation of an equation of motion for the slowly time-de\-pen\-dent
den\-si\-ty operator $\hat{\rho}_P(t)$ and introduction of perturbation- and
Markov- approximations}    \label{derivation}  
\setcounter{equation}{0}

The theory to be presented in this paper is devised to describe the slowly
time-dependent average evolution of a fission-- or fusion--process at
moderate excitation energies. It should thus be applicable to reactions which
proceed via the formation of compound nuclear resonances (``CNR") with a
lifetime which is much larger than the time scale of its formation
($\lesssim$ 10$^{-22}$ s). The CNR are highly complicated nuclear many-body
states which exhibit a finite width due to their coupling to the continua of
the open decay channels. It is impossible to determine the detailed form of
these states. According to the deep original insight of N. Bohr
\cite{NBohr}, the detailed structure of the CNRs is not
important for the understanding of the slow time-evolution of the system in
the fission channel which is the topic of the present paper.
In the language of transport theory, Bohr's hypothesis implies that the
``memory time" of the system is much smaller than the time-scale of the
collective motion in the fission channel. Consequently, we may describe the
slow collective time-evolution of the fission process starting from a simple
initial condition for the collective flow. We shall come back to this
question when discussing the ``transport equation" we are going to derive.

One usually considers the decay of an ensemble of CN resonances. But even in
the case that we study the decay of a single, isolated CN resonance, the
system should be described by the ``density operator" (=``statistical
operator") $\hat{\rho}(t)$  rather than by a wavefunction. The reason is that
it is the density operator which expresses the information we have on a
physical system and it is this quantity which simplifies when we ask for a
reduced information only.

The statistical operator satisfies the von Neumann-Liouville equation
\begin{equation}\label{3.1}
\dot{\hat{\rho}}(t) = \frac{-i}{\hbar} \left[\hat{H},\hat{\rho}(t)\right]
\end{equation}
In our case, $\hat{H}$ is the Hamiltonian of $A$ nucleons ($Z$ protons and
$N$ neutrons) interacting by effective nucleon-nucleon interactions $v$ :
\begin{equation}\label{3.2}
\hat{H} =-\sum_{i=1}^{A} \frac{\hbar^2}{2M} \Delta(i)
+\frac{1}{2} \sum_{i,j=1}^{A} v(x_i,x_j)
\end{equation}
The symbol $x_i$ denotes the position, spin and isospin variables of the
$i$-th nucleon and $M$ is the mass of the nucleon. We emphasize that the
knowledge of the nucleon-nucleon interaction represents the only nuclear
input information in our theory.

The statistical operator $\hat{\rho}(t)$ describes the entire time evolution,
i.e. the slow collective change of the nucleonic distribution during fission
as well as the rapidly changing state of internal motion which involves a
superposition of many complicated intrinsic excitation modes. Of course, we
also need a knowledge of the initial state $\hat{\rho}(0)$. Due to the effect
of memory loss mentioned above, it is probably innocuous that we only have a
rather incomplete information on $\hat{\rho}(0)$.

At the origin of any transport theory, one has to introduce the
part $\hat{\rho}_P(t)$ of the total density operator which should describe
the slowly proceeding dynamical evolution of the system only. As we do not
have independent collective degrees of freedom at our disposal, we have to
define the density operator $\hat{\rho}_P(t)$ in terms of an ``ansatz" for
the matrix-elements
$\langle a_1 q_1\vert \hat{\rho}_P(t) \vert a_2 q_2\rangle$
of $\hat{\rho}_P(t)$      with respect to the basis-states introduced in
Section 2.\\
As the slow collective time change is not expected to depend on the precise
form of the intrinsic motion, let us first define a ``reduced density matrix"
\begin{equation}\label{3.3}
R(q_1, q_2 ; t) \equiv \rho^{red} (q_1, q_2 ; t) := \sum_{a}
\langle a q_1\vert \hat{\rho}(t) \vert a q_2\rangle
\end{equation}
as the part of the operator $\hat{\rho}_P(t)$ which is to describe the
dynamics of the shape changes and, thereafter, of the motion of the
fission fragments in space.\\
Furthermore, we introduce a grand canonical distribution of the system with
regard to the eigen-energies ${\cal E}_a(q)$ of the basis states
\begin{equation}\label{3.4}
F_a(q,T(q,t)):=\frac{1}{Z_0(q,T)} \exp{\left(-\beta(q,t){\cal E}'_a(q)\right)}
\end{equation}
Here, $\beta$ is the reciprocal temperature
\begin{equation}\label{3.5}
\beta(q,T)=\frac{1}{T(q,t)}
\end{equation}
and $Z_0(q,T)$  the partition function
\begin{equation}\label{3.6}
Z_0(q,T):=\sum_{a} e^{-\beta(q,t){\cal E}'_a(q)}
\end{equation}
The temperature $T$ may depend on the g.c. $q$ and on the time $t$. The
choice of the temperature $T(q,t)$ as a function of $q$ and $t$ will be
presented in Section 4. Let us only note that, for the case of a ``nuclear
transport theory", the ``temperature" is not determined externally by the
contact of the system with a bath, i.e. a large reservoir at a given
temperature. Rather, we must consider it as a parameter which we have to
choose so as to optimize the ``ansatz" $\hat{\rho}_P(t)$ for the
slowly time-dependent part of the density operator $\hat{\rho}(t)$.

We define the matrix representation of $\hat{\rho}_P(t)$ by
\begin{equation}\label{3.7}
\langle a_1 q_1\vert \hat{\rho}_P(t) \vert a_2 q_2 \rangle :=
\delta_{a_1a_2} R(q_1,q_2;t) F_{a_2}\left(q_2,T(q_2,t)\right)
\end{equation}
The Boltzmann distribution (\ref{3.4}) which appears in the ansatz
(\ref{3.7}) is customarily also written in the form
\begin{equation}\label{3.8}
F_a(q,T(q,t))=\exp{\left[-\beta(q,t)\left({\cal E}'_a(q)
-G(q,T)\right)\right]}
\end{equation}
where the Gibbs function $G(q,T)$ is related to the partition function
(\ref{3.6}) by
\begin{equation}\label{3.9}
G(q,T)= -T(q,t)\ln{Z_0(q,T)}
\end{equation}
The following idea underlies the choice of the canonical distribution of the
system with respect to the intrinsic excitations in Eq. (\ref{3.7}):\\
As $\hat{\rho}_P(t)$ should only describe the slow time-evolution of the
system, the precise state of the intrinsic motion is not believed to be
relevant. The choice of the canonical distribution as approximation may be
motivated by the observation that the distribution corresponds to the minimal
information (=maximal entropy) on the state of internal motion (see Section
4).

We draw attention to the fact that the matrix (\ref{3.7}) does not define a
hermitian operator $\hat{\rho}_P(t)$,
even if we use orthonormal basis states $ \vert a q \rangle$.
In addition, let us note that the reduced matrix $R(q_1, q_2 ; t)$ is
hermitian only if, in Eq. (\ref{3.3}), we use an orthonormal basis.
It is not hermitian if we use the biorthogonal basis.
A further discussion of the question of hermiticity will be given in the
Appendix \ref{appendix3}.

The non-hermitian character of $\hat{\rho}_P(t)$ is of no consequence as far
as mean-values of observables $\hat{O}$=$\hat{O}^\dagger$ are concerned. We
just have to evaluate the mean values with the hermitian part of
$\hat{\rho}_P(t)$ which is defined by
\begin{equation}\label{3.10}
\hat{\rho}^h_P(t)=\frac{1}{2}\left(\hat{\rho}_P(t)+
\hat{\rho}^\dagger_P(t)\right)
\end{equation}
We find
\begin{align}
\langle \langle \hat{O} \rangle\rangle :=&
{\rm Tr}\left( \hat{\rho}^h_P(t) \hat{O} \right)
= \sum_{a_1q_1a_2q_2} {\rm Re}\left\{
\langle a_1q_1\vert \hat{\rho}_P(t)\vert a_2q_2\rangle
\langle a_2q_2\vert \hat{O}\vert a_1q_1\rangle
\right\}  \notag\\
=& {\rm Re}\left\{{\rm  Tr}\left(\hat{\rho}_P(t) \hat{O}\right)\right\}
\label{3.11}
\end{align}
We also note that, for orthonormal basis states $ \Psi_a(q)$, we can
interpret
\begin{equation}\label{3.12}
R(q,q;t) := \sum_{a} \langle \Psi_a(q)\vert\hat{\rho}(t) \vert
\Psi_a(q)\rangle
\end{equation}
as the probability to find the system in any one of the basis states
$\Psi_a(q)$ for given $q$. In the case of the biorthogonal basis, the
corresponding quantity
\begin{equation}\label{3.13}
R(q,q;t) := \sum_{a} \langle \widetilde{\phi_a(q)} \vert\hat{\rho}(t) \vert
\phi_a(q)\rangle
\end{equation}
cannot be interpreted as a probability, as
$\vert \widetilde{\phi_a(q)} \rangle\langle \phi_a(q)\vert$
is not a hermitian projection operator contrary to
$\vert \Psi_a(q) \rangle\langle\Psi_a(q)\vert$.

We can write the operator $\hat{\rho}_P(t)$ as a projection operator
$\hat{P}$ acting on $\hat{\rho}(t)$:
\begin{equation}\label{3.14}
 \hat{\rho}_P(t) =  \hat{P} \hat{\rho}(t)
\end{equation}
Writing $\langle a_1q_1\vert  \hat{\rho}_P(t) \vert a_2q_2 \rangle$
as defined by Eq. (\ref{3.7}) in the form
\begin{equation}\label{3.15}
\langle a_1q_1\vert  \hat{P} \hat{\rho}(t) \vert a_2q_2 \rangle
=\sum_{a_3q_3a_4q_4} P_{a_1q_1a_2q_2a_3q_3a_4q_4}
\langle a_3q_3\vert  \hat{\rho}(t) \vert a_4q_4 \rangle
\end{equation}
one finds that the matrix representation of the projection operator
$\hat{P}$ has the form
\begin{equation}\label{3.16}
P_{a_1q_1a_2q_2a_3q_3a_4q_4} =
\delta_{a_1a_2} \delta_{q_1q_3} \delta_{q_2q_4} \delta_{a_3a_4}
F_{a_2}(q_2,T(q_2,t))
\end{equation}
Using the definition (\ref{3.16}) of  $\hat{P}$ we define a complementary
projection operator
\begin{equation}\label{3.17}
\hat{Q} =1- \hat{P}
\end{equation}
by the matrix representation
\begin{equation}\label{3.18}
Q_{a_1q_1a_2q_2a_3q_3a_4q_4} =
\delta_{a_1a_3} \delta_{q_1q_3} \delta_{a_2a_4} \delta_{q_2q_4} -
P_{a_1q_1a_2q_2a_3q_3a_4q_4}
\end{equation}
One easily checks that the following relations hold:
\begin{equation}\label{3.19}
\hat{P}^2 =\hat{P} ;\hspace{5mm} \hat{Q}^2 =\hat{Q}
 ;\hspace{5mm} \hat{P}\hat{Q} =\hat{Q}\hat{P}=0
\end{equation}
Through their dependence on the (slowly) time-dependent temperature, the
projection operators are slowly time-dependent quantities.
The total density operator $\hat{\rho}(t)$ can be decomposed into the slowly
time-dependent part $\hat{\rho}_P(t)$ and the remainder $\hat{\rho}_Q(t)$:
\begin{equation}\label{3.20}
\hat{\rho}(t) =\left(\hat{P}+\hat{Q}\right)\hat{\rho}(t)=
\hat{\rho}_P(t) +\hat{\rho}_Q(t)
\end{equation}
The part $\hat{\rho}_Q(t)$ of the statistical operator describes rapidly
time-dependent processes and involves summations over a large number of terms
with different phases. Whereas the detailed form of $\hat{\rho}_Q(t)$ will be
considered in the next Section, let us note the properties
\begin{align}
& \sum_{a}\langle a  q_1\vert \hat{\rho}_Q(t)\vert a  q_2\rangle=0
\label{3.21} \\
& {\rm Tr} \left(\hat{\rho}_Q(t) \hat{O}^{red}\right) =0
\label{3.22}
\end{align}
Here, $\hat{O}^{red}$ is an operator the matrix-representation of which is
diagonal with respect to the quantum numbers of the intrinsic motion:
\begin{equation}\label{3.23}
\langle a_2q_2\vert \hat{O}^{red}\vert a_1q_1\rangle
:=\delta_{a_2a_1} O(q_2,q_1)
\end{equation}
Obviously, observables with the property (\ref{3.22}) put to the test purely
collective features of the system.
Examples are the charge-- and mass-- numbers and the kinetic energy of
relative motion of the fission fragments.

For deriving an equation of motion for the operator $\hat{\rho}_P(t)$, we
make use of the formalism introduced by Nakajima \cite{Nakajima} and
Zwanzig \cite{Zwanzig}. Acting on the Eq. (\ref{3.1}) with the projection
operators $\hat{P}$ and $\hat{Q}$ and taking into account their
time-dependence, we obtain the coupled equations
\begin{align}
\frac{d\hat{\rho}_P(t)}{dt}=& -\frac{i}{\hbar} \hat{P}\hat{L}
\left(\hat{\rho}_P(t)+\hat{\rho}_Q(t)\right)
+ \dot{\hat{P}}(t) \hat{\rho}(t) \label{3.24}\\
\frac{d\hat{\rho}_Q(t)}{dt}=& -\frac{i}{\hbar} \hat{Q}\hat{L}
\left(\hat{\rho}_P(t)+\hat{\rho}_Q(t)\right)
+ \dot{\hat{Q}}(t) \hat{\rho}(t) \label{3.25}
\end{align}
where the Liouvillean operator $\hat{L}$  acting on an operator $\hat{A}$ is
defined by
\begin{equation}\label{3.26}
\hat{L}\hat{A} := \left[\hat{H},\hat{A}\right]
\end{equation}
Using some properties of projection operators and their time derivatives,
which are derived and presented in Appendix \ref{appendix1}, we may write
\begin{equation}\label{3.27}
\dot{\hat{P}}(t) \hat{\rho}(t) =
\dot{\hat{P}}(t)\left(\hat{P}(t)+\hat{Q}(t)\right) \hat{\rho}(t)
= \dot{\hat{P}}(t)\hat{\rho}_P(t)
\end{equation}
and
\begin{equation}\label{3.28}
\dot{\hat{Q}}(t) \hat{\rho}(t) = - \dot{\hat{P}}(t) \hat{\rho}(t)
= - \dot{\hat{P}}(t)\hat{\rho}_P(t)
\end{equation}
The coupled equations (\ref{3.24})-(\ref{3.25}) thus take the form
\begin{align}
\frac{d\hat{\rho}_P}{dt}=& -\frac{i}{\hbar} \hat{P}\hat{L}
\left(\hat{\rho}_P(t)+\hat{\rho}_Q(t)\right)
+\dot{\hat{P}}(t)\hat{\rho}_P(t) \label{3.29}\\
\frac{d\hat{\rho}_Q}{dt}=& -\frac{i}{\hbar} \hat{Q}\hat{L}
\hat{\rho}_Q(t)+\hat{\sigma}(t)         \label{3.30}
\end{align}
where the ``source term" $\hat{\sigma}(t)$ is defined by
\begin{equation}\label{3.31}
\hat{\sigma}(t) := -\frac{i}{\hbar} \hat{Q}(t)\hat{L}\hat{\rho}_P(t)
- \dot{\hat{P}}(t)\hat{\rho}_P(t)
\end{equation}
In order to obtain an equation of motion for the density operator
$\hat{\rho}_P(t)$, we have to find a formal solution of the Eq. (\ref{3.30})
and substitute it into Eq. (\ref{3.29}).
Proceeding in analogy to Ref. \cite{Balian}, we introduce a Green function
$\hat{G}(t,s)$ as a solution of the equation
\begin{equation}\label{3.32}
\frac{\partial \hat{G}(t,s)}{\partial t}+\frac{i}{\hbar}
\hat{Q}(t)\hat{L}\hat{G}(t,s)
=\hat{Q}(t)\delta(t-s)
\end{equation}
and write the operator $\hat{\rho}_Q(t)$ for $t>0$ in the form
\begin{equation}\label{3.33}
\hat{\rho}_Q(t) =\hat{G}(t,0) \hat{\rho}_Q(0) +
\int_0^{\infty} ds \,\hat{G}(t,s)  \hat{Q}(s) \hat{\sigma}(s)
\end{equation}
In the Appendix \ref{appendix1}, it is shown that the operator
$\dot{\hat{P}}(t)$  has the property
\begin{equation}\label{3.34}
\hat{P}(t) \dot{\hat{P}}(t) =0
\end{equation}
which implies
\begin{equation}\label{3.35}
\dot{\hat{P}}(t) =\left(\hat{P}(t) +\hat{Q}(t) \right) \dot{\hat{P}}(t)
= \hat{Q}(t)\dot{\hat{P}}(t)
\end{equation}
Consequently, the source term $\hat{\sigma}(s)$ satisfies
\begin{equation}\label{3.36}
\hat{Q}(s)\hat{\sigma}(s)= \hat{\sigma}(s)
\end{equation}
and we may write, instead of Eq. (\ref{3.33}), more simply
\begin{equation}\label{3.37}
\hat{\rho}_Q(t) =\hat{G}(t,0) \hat{\rho}_Q(0) +
\int_0^{\infty} ds \,\hat{G}(t,s)  \hat{\sigma}(s)
\end{equation}
It is easily checked that the ansatz (\ref{3.37}) together with the defining
equation (\ref{3.32}) for the Green operator $\hat{G}(t,s)$ represents the
solution of the Eq. (\ref{3.30}) with the initial value $\hat{\rho}_Q(0)$:
\begin{align}
\frac{d\hat{\rho}_Q(t)}{dt} \stackrel{(\ref{3.37})}{=}&
\frac{\partial \hat{G}(t,0)}{\partial t} \hat{\rho}_Q(0) +
\int_0^{\infty} ds \frac{\partial \hat{G}(t,s)}{\partial t}  \hat{\sigma}(s)
\notag\\
\stackrel{(\ref{3.32})}{=}& \left[
  -\frac{i}{\hbar} \hat{Q}(t)\hat{L}\hat{G}(t,0) +\hat{Q}(t)\delta(t)
  \right]  \hat{\rho}_Q(0)
+ \int_0^{\infty} ds \left[
       -\frac{i}{\hbar} \hat{Q}(t)\hat{L}\hat{G}(t,s) +\hat{Q}(t)\delta(t-s)
\right]\hat{\sigma}(s) \notag\\
\frac{d\hat{\rho}_Q(t)}{dt}  \stackrel{(\ref{3.37})}{=}&
-\frac{i}{\hbar} \hat{Q}(t)\hat{L}\hat{G}(t,0)\hat{\rho}_Q(0)
  +\hat{Q}(0) \hat{\rho}_Q(0) \delta(t)
-\frac{i}{\hbar} \hat{Q}(t)\hat{L} \left[
  \hat{\rho}_Q(t) -\hat{G}(t,0) \hat{\rho}_Q(0)   \right]
  +  \theta_0(t) \hat{Q}(t) \hat{\sigma}(t)
\label{3.38}
\end{align}
Apart from the term $\hat{Q}(0) \hat{\rho}_Q(0) \delta(t)$, which expresses
the initial condition for $\hat{\rho}_Q(t)$, the r.h.s. of Eq. (\ref{3.38})
is seen to agree for $t>0$ with the r.h.s. of Eq. (\ref{3.30}).

Substituting the formal solution (\ref{3.37}) into the Eq. (\ref{3.29}), we
obtain the desired integro-differential equation for $\hat{\rho}_P(t)$:
\begin{align}
\frac{d\hat{\rho}_P(t)}{dt}=&
 -\frac{i}{\hbar} \hat{P}(t)\hat{L}\hat{G}(t,0) \hat{\rho}_Q(0)
 -\frac{i}{\hbar} \hat{P}(t)\hat{L} \hat{\rho}_P(t) \notag\\
 &-\frac{i}{\hbar} \hat{P}(t)\hat{L}
 \int_0^{\infty} ds \,\hat{G}(t,s)  \hat{\sigma}(s)
+\dot{\hat{P}}(t)\hat{\rho}_P(t) \label{3.39}
\end{align}
We will refer to this equation as ``Nakajima-Zwanzig (NZ)"-equation.
It is a rigorously valid equation of motion for the slowly time-dependent
part $\hat{\rho}_P(t)$ of the total density operator $\hat{\rho}(t)$.

The formal solution of the equation (\ref{3.32}) for the Green operator
$\hat{G}(t,s)$ can be written as the following time-ordered product
\begin{equation}\label{3.40}
\hat{G}(t,s)=\theta_0(t-s)  \hat{Q}(t)  \hat{T}\left\{e^{\textstyle
-\int_s^t d\tau
\left[\frac{i}{\hbar}\hat{L}\hat{Q}(\tau)-\dot{\hat{P}}(\tau)\right]}\right\},
\end{equation}
$\hat{T}$ denoting the time-ordering operator.
The detailed proof that (\ref{3.40}) is a solution of Eq. (\ref{3.32}) is
given in the Appendix \ref{appendix2}.
Through its dependence on the Green operator $\hat{G}(t,s)$, the NZ-equation
(\ref{3.39}) still contains the full complexity of the total system.
Nevertheless, this equation has the merit to display the different physical
processes which contribute to the time evolution of $\hat{\rho}_P(t)$.

In order to describe the physical meaning of the different terms in
the equation of motion (\ref{3.39}), it is useful to decompose the matrix
$\langle a_1q_1\vert \hat{A}\vert a_2q_2\rangle$
representing an arbitrary operator $\hat{A}$ into two parts using the
projection operators $\hat{P}$ and $\hat{Q}$:
\begin{align}
&  \langle a_1q_1\vert \hat{A}\vert a_2q_2\rangle
=\langle a_1q_1\vert\hat{P} \hat{A}\vert a_2q_2\rangle
+\langle a_1q_1\vert\hat{Q} \hat{A}\vert a_2q_2\rangle \label{3.41}\\
& \langle a_1q_1\vert\hat{P} \hat{A}\vert a_2q_2\rangle
 =\delta_{a_1a_2} F_{a_2}\left(q_2,T(q_2,t)\right) A^{red}(q_1,q_2)
\label{3.42}\\
& A^{red}(q_1,q_2):=\sum_{a} \langle a  q_1\vert \hat{A}\vert a  q_2\rangle
\label{3.43}\\
&  \langle a_1q_1\vert\hat{Q} \hat{A}\vert a_2q_2\rangle :=
 \langle a_1q_1\vert \hat{A}\vert a_2q_2\rangle
 - \delta_{a_1a_2} F_{a_2}(q_2,T(q_2,t)) A^{red}(q_1,q_2)  \label{3.44}
\end{align}
We shall refer to the two parts
$\langle a_1q_1\vert\hat{P} \hat{A}\vert a_2q_2\rangle$ and
$\langle a_1q_1\vert\hat{Q} \hat{A}\vert a_2q_2\rangle$ as the ``canonical"
or ``$P$-part" and the ``non-canonical" or ``$Q$-part" of the matrix
$\langle a_1q_1\vert \hat{A}\vert a_2q_2\rangle$.\\
Obviously, the Eq. (\ref{3.41}) generalizes the decomposition of the density
operator $\hat{\rho}$ to one of an arbitrary operator $\hat{A}$.
Whereas  $\langle a_1q_1\vert\hat{P} \hat{A}\vert a_2q_2\rangle$ depends on
intrinsic excitations  only through a canonical distribution,
the complementary part $\langle a_1q_1\vert\hat{Q} \hat{A}\vert a_2q_2\rangle$
is expected to depend on the quantum numbers $a_1$ and $a_2$ in a
complicated, possibly almost random way.

The nature of the different terms in the NZ-equation (\ref{3.39}) can be
described as follows:\\
Except for the term $\dot{\hat{P}}(t)\hat{\rho}_P(t)$, all terms on the
r.h.s. of the NZ-equation (\ref{3.39}) contain the projection
operator $\hat{P}(t)$ on their left, which means that their matrix
representation is of the form (\ref{3.42}), (\ref{3.43}).
The term $\dot{\hat{P}}(t)\hat{\rho}_P(t)$ has the matrix representation
\begin{align}
\langle a_1q_1\vert\dot{\hat{P}}(t)\hat{\rho}_P(t) \vert a_2q_2\rangle=&
\sum_{a_3q_3a_4q_4} \dot{P}_{a_1q_1a_2q_2a_3q_3a_4q_4}
\langle a_3q_3\vert\hat{\rho}_P(t) \vert a_4q_4\rangle\notag\\
=& \delta_{a_1a_2}  \frac{\partial F_{a_2}(q_2,T(q_2,t))}{\partial  T}
 \dot{T}(q_2,t) R(q_1,q_2;t)
\label{3.45}
\end{align}
where the derivative $\partial F_{a_2}/\partial T$ can be
written in the form
\begin{align}
& \frac{\partial F_{a_2}(q_2,T(q_2,t))}{\partial T}
=\frac{F_{a_2}(q_2,T(q_2,t))}{T}\left\{
\frac{{\cal E}'_{a_2}(q_2)-G(q_2,T)}{T} +
\frac{\partial G(q_2,T)}{\partial T} \right\}
\label{3.46}
\end{align}
Introducing the entropy $S(q_2,T)$ of the system at the given value $q_2$ of
the g.c.
\begin{equation}\label{3.47}
S(q_2,T)= -\frac{\partial G(q_2,T)}{\partial T}
\end{equation}
we can present the term
$\langle a_1q_1\vert\dot{\hat{P}}(t)\hat{\rho}_P(t) \vert a_2q_2\rangle$
as follows
\begin{align}
\langle a_1q_1\vert\dot{\hat{P}}(t)\hat{\rho}_P(t) \vert a_2q_2\rangle
=&  \delta_{a_1a_2}  R(q_1,q_2;t)  F_{a_2}(q_2,T(q_2,t))
\frac{\dot{T}(q_1,t)}{T(q_2,t)}
\left\{ \frac{{\cal E}'_{a_2}(q_2)-G(q_2,T)}{T} -S(q_2,T(q_2,t)) \right\}
\label{3.48}
\end{align}
As the temperature is expected to be a slow function of time, the term is
likely to be smaller than the other terms on the r.h.s of the NZ-equation.
The role of the entropy will be discussed in Section \ref{determination}.\\
The inhomogeneous term
$-i/\hbar\, \hat{P}(t)\hat{L}\hat{G}(t,0) \hat{\rho}_Q(0)$
represents the memory of the system of the ``non-canonical" part
$\hat{\rho}_Q(0)$ of the initial density operator $\hat{\rho}(0)$.
The operator $\hat{G}(t,0)$  propagates $\hat{\rho}(0)$ from time 0 to time
$t$, so that $-i/\hbar\, \hat{L}\hat{G}(t,0)
\hat{\rho}_Q(0)$   essentially represents the time-derivative of
$\hat{G}(t,0) \hat{\rho}_Q(0)$. Finally, the operator $ \hat{P}(t)$  projects
from it the component which is of $P$-type. One expects that the term
$-i/\hbar\, \hat{P}(t)\hat{L}\hat{G}(t,0) \hat{\rho}_Q(0)$
fades away after a short ``memory time" $\tau_{mem} \leq 10^{-22}$ s. In
fact, the projection operator $\hat{P}$ filters from
$\langle a_1q_1\vert\hat{L}\hat{G}(t,0) \hat{\rho}_Q(0) \vert a_2q_2\rangle$
the ``reduced part"
\begin{equation}\label{3.49}
\langle a_1q_1\vert\hat{P}(t)\hat{L}\hat{G}(t,0) \hat{\rho}_Q(0)
\vert a_2q_1\rangle =\delta_{a_1a_2} F_{a_2}(q_2,T)
\sum_{a} \langle a q_1\vert\hat{L}\hat{G}(t,0) \hat{\rho}_Q(0)
\vert a q_2\rangle
\end{equation}
As the density operator $\hat{\rho}_Q(0)$ represents the complicated part of
the initial density $\hat{\rho}(0)$ and as the propagator $\hat{G}(t,0)$
tends to complicate further this part of the statistical operator, the
$\hat{P}$-projection (\ref{3.49}) is expected to become rapidly negligible.\\
In all what follows, the inhomogeneous term
$-i/\hbar\, \hat{P}(t)\hat{L}\hat{G}(t,0) \hat{\rho}_Q(0)$
of the NZ-equation will be neglected. The term was also neglected in all
actual calculations of the fission dynamics we know of
\cite{Hofmann,Berger-Gogny,Goutte,KDKP}.

Let us note that neglecting this term means essentially that one chooses
the initial condition at a time where the collective motion towards fission
has already started. The original compound system consists, at low excitation
energy ($\lesssim$ 8 MeV), of several narrow ($\simeq$ 1 eV) compound
nuclear resonances and, at high excitation energy ($\gtrsim$ 10 MeV), of
overlapping resonances of an average width of $\gtrsim$ 100 keV. Thus, just
after its formation, the compound system does not at all resemble a canonical
distribution over intrinsic excitations as required by $\hat{\rho}_P(t)$, but
rather a microcanonical ensemble of highly complex nuclear decaying states.
The dynamical evolution of this initial state into one whose slowly
time-dependent part is approximately canonically distributed has so far never
been studied in detail.

The terms $-i/\hbar \hat{P}(t)\hat{L} \hat{\rho}_P(t)$
and $\ds -i/\hbar \hat{P}(t)\hat{L} \int_0^{\infty} ds \ldots $
in the NZ-equation (\ref{3.39}) can be physically interpreted as follows:\\
As we can see from Eq. (\ref{3.29}), the term
$-i/\hbar \hat{P}(t)\hat{L} \hat{\rho}_P(t)$ would yield the whole
time-derivative $d\hat{\rho}_P(t)/dt$ if, at time $t$, the state of the
system $\hat{\rho}(t)$ were entirely given by $\hat{\rho}_P(t)$. However,
there is also the part described by $\hat{\rho}_Q(t)$, and thus there is the
contribution
$-i/\hbar \hat{P}(t)\hat{L} \hat{\rho}_Q(t)$ to the time-rate
$d\hat{\rho}_P(t)/dt$.
As has been shown, this term takes the form
$\ds -i/\hbar \hat{P}(t)\hat{L} \int_0^{\infty} ds
\hat{G}(t,s)  \hat{\sigma}(s)$.
It describes that part of the rapid complicated time evolution which
contributes to the change of the slowly time-dependent density operator
$\hat{\rho}_P(t)$. As we shall see that this term
describes friction and dissipation processes, we shall refer to it as the
``dissipation term" of the NZ-equation.

As it stands, the NZ-equation (\ref{3.39}) cannot be solved due to the
complexity of the propagator $\hat{G}(t,s)$. As this propagator appears in
the equation of motion of the slowly time-dependent density operator
$\hat{\rho}_P(t)$, one may introduce some substantial simplifications, namely
perturbation theory within the dissipative term and the Markov approximation.

Let us first define a ``collective potential Hamiltonian" $\hat{H}_{coll}$ as
a canonical average of the potential surfaces which correspond to the
different intrinsic excitations of the system:\\
In Eq. (\ref{3.50}), we first introduce a Hamiltonian $\hat{H}_1$ with the
same eigenstates as the basis Hamiltonian $\hat{H}'_1$ or $\hat{{\cal H}}'$
resp., but with shifted eigenvalues
\begin{equation}\label{3.50}
\hat{H}_1 :=\sum_{aq} \vert aq\rangle \left({\cal E}'_a(q)+\lambda(q)
\langle aq\vert \hat{q}\vert aq\rangle\right) \langle aq \vert
\end{equation}
or, separately for the biorthogonal and the orthonormal basis, resp.,
\begin{align}
& \hat{H}_1 = \left\{\begin{array}{l} \ds
\sum_{aq} \vert \phi_a(q) \rangle \left({E'}_{a}^{(1)}(q)+\lambda(q)
\langle \widetilde{\phi_a(q)}\vert \hat{q}\vert \phi_a(q)\rangle\right)
\langle \widetilde{\phi_a(q)} \vert \\\ds
\sum_{aq} \vert \Psi_a(q)\rangle \left({\varepsilon}'_a(q)+\lambda(q)
\langle \Psi_a(q)\vert \hat{q}\vert \Psi_a(q)\rangle\right)
\langle \Psi_a(q) \vert
\end{array} \right.
\label{3.50'} \tag{3.50'}
\end{align}
\begin{equation}\label{3.51}
\hat{H}_1\vert aq\rangle= \left({\cal E}'_a(q)
+\lambda(q) \langle aq\vert \hat{q}\vert aq\rangle\right)
\vert aq\rangle
\end{equation}
The reason for shifting the eigenvalues by the amount
$\lambda(q) \langle aq\vert \hat{q}\vert aq\rangle$ is that $\hat{H}_1$
should be a simple part of the total Hamiltonian $\hat{H}$ rather than of the
Routhian $\hat{H}'=\hat{H}-\lambda\hat{q}$. We denote the new eigen-energies
by
\begin{equation}\label{3.52}
{\cal E}_a(q) :=  {\cal E}'_a(q)
+\lambda(q) \langle aq\vert \hat{q}\vert aq\rangle
\end{equation}
The matrix-representation of $\hat{H}_{coll}$ is then defined as
\begin{equation}\label{3.53}
\langle a_1q_1\vert \hat{H}_{coll}\vert a_2q_2\rangle
=\delta_{a_1a_2} \delta_{q_1q_2} \sum_{a} {\cal E}_a(q_2)F_a(q_2,T(q_2,t))
\end{equation}
i.e. as the canonical mean value of the Hamiltonian $\hat{H}_1$.
Due to the time-dependence of the temperature, the matrix
$\langle a_1q_1\vert \hat{H}_{coll}\vert a_2q_2\rangle$
acquires a slow dependence on time.

The eigen-energies ${\cal E}_a(q)$  represent a family of potential surfaces,
one for each quantum state $a$ of the intrinsic motion. As we have discussed
in Section 2, the quantum number ``$a$" and the generator coordinate $q$
design a particular HW-state $\phi_a(q)$, if the biorthogonal basis is
used, and they denote the largest component $\phi_a(q)$ in the basis state
$\Psi_a(q)$, if the orthonormal basis is used.\\
In the limit of vanishing temperature, the canonical mean value of
the different potential surfaces becomes equal to the
lowest potential surface ${\cal E}_{a_0}(q)$:
\begin{equation}\label{3.54}
{\protect \parbox[t]{15mm}{\hspace{2mm}{\em   lim} \\
          \raisebox{2mm}{$\scriptstyle T(q_2,t)\rightarrow 0$}}}
\sum_{a} {\cal E}_a(q)F_a(q,T) ={\cal E}_{a_0}(q)
\end{equation}
We define the coupling Hamiltonian $\hat{H}_{cpl}$ as the difference between
the total Hamiltonian $\hat{H}$ and $\hat{H}_{coll}$:
\begin{equation}\label{3.55}
\hat{H}_{cpl}=\hat{H}-\hat{H}_{coll}
\end{equation}
with the matrix-representation
\begin{equation}\label{3.56}
\langle a_1q_1\vert \hat{H}_{cpl}\vert a_2q_2\rangle=
\langle a_1q_1\vert \hat{H}\vert a_2q_2\rangle-
\langle a_1q_1\vert \hat{H}_{coll}\vert a_2q_2\rangle
\end{equation}
Using the Wick decomposition for $\hat{H}'$, we obtain a more explicit form
of $\langle a_1q_1\vert \hat{H}_{cpl}\vert a_2q_2\rangle$:
\begin{align*}
\langle a_1q_1\vert \hat{H}_{cpl}\vert a_2q_2\rangle = &
\langle a_1q_1\vert \hat{H}-\lambda(q_2)\hat{q}\vert a_2q_2\rangle+
\lambda(q_2)\langle a_1q_1\vert \hat{q}\vert a_2q_2\rangle
- \langle a_1q_1\vert \hat{H}_{coll}\vert a_2q_2\rangle
\end{align*}
or
\begin{align}
\langle a_1q_1\vert \hat{H}_{cpl}\vert a_2q_2\rangle =&
\delta_{a_1a_2}\delta_{q_1q_2}
\left({\cal E}'_{a_2}(q_2) -\langle \langle {\cal E}'(q_2) \rangle\rangle \right)
+\lambda(q_2)\left(\langle a_1q_1\vert \hat{q}\vert a_2q_2\rangle
-  \langle \langle \hat{q}(q_2) \rangle \rangle\right)
+\langle a_1q_1\vert \hat{H}^{'^{(4)}}_2(q_2)\vert a_2q_2\rangle
\label{3.57}
\end{align}
Here, we have introduced a short notation for canonical mean values
\begin{align}
& \langle \langle {\cal E}'(q) \rangle\rangle = \sum_{a}
{\cal E}'_a(q)  F_a(q,T(q,t))                        \label{3.58}  \\
&\langle \langle \hat{q}(q) \rangle \rangle =\sum_{a}
\langle aq\vert \hat{q}\vert aq\rangle F_a(q,T(q,t))  \label{3.59}
\end{align}
The first term in Eq. (\ref{3.57}) represents the fluctuation of the
intrinsic energies around their thermal mean value at a given value of the
g.c. $q_2$, and the second term the fluctuation of the ``collective transport
term" $\lambda(q_2)\langle a_1q_1\vert \hat{q}\vert a_2q_2\rangle$ around its
thermal mean value at the g.c. $q_2$.
The last term in Eq. (\ref{3.57}) is given by the matrix-element of the
2-body part $\hat{H}^{'^{(4)}}_2(q_2)$ of the Hamiltonian $\hat{H}'(q_2)$.
The total matrix $\langle a_1q_1\vert \hat{H}_{cpl}\vert a_2q_2\rangle$
is hermitian if we use the orthonormal basis functions $\Psi_a(q)$:
\begin{align*}
\langle \Psi_{a_1}(q_1)\vert \hat{H}_{cpl}\vert \Psi_{a_2}(q_2)\rangle=
\langle \Psi_{a_2}(q_2)\vert \hat{H}_{cpl}\vert \Psi_{a_1}(q_1)\rangle^*
\end{align*}
whereas the different terms on the r.h.s. of (\ref{3.57}) are not hermitian.
This is due to the fact that the Routhian
$\hat{H}'(q)=\hat{H}-\lambda(q)\hat{q}$
depends on the value of the generator coordinate.\\
One can easily see that the Liouvillean $\hat{L}_{coll}$ associated with
$\hat{H}_{coll}$ commutes with the projection operators: The matrix
representation of the Liouvilleans read:
\begin{align}
& L_{a_1q_1a_2q_2a_3q_3a_4q_4} = \delta_{a_2a_4} \delta_{q_2q_4}
\langle a_1q_1\vert \hat{H}\vert a_3q_3\rangle
-\delta_{a_1a_3} \delta_{q_1q_3}\langle a_4q_4\vert \hat{H}\vert a_2q_2\rangle
\label{3.60}\\
&{L_{cpl}}_{a_1q_1a_2q_2a_3q_3a_4q_4} = \delta_{a_2a_4} \delta_{q_2q_4}
\langle a_1q_1\vert \hat{H}_{cpl}\vert a_3q_3\rangle
-\delta_{a_1a_3} \delta_{q_1q_3}\langle a_4q_4\vert \hat{H}_{cpl}
\vert a_2q_2\rangle
\label{3.61}\\
&{L_{coll}}_{a_1q_1a_2q_2a_3q_3a_4q_4} : = \delta_{a_2a_4} \delta_{q_2q_4}
\langle a_1q_1\vert \hat{H}_{coll}\vert a_3q_3\rangle
-\delta_{a_1a_3} \delta_{q_1q_3}\langle a_4q_4\vert \hat{H}_{coll}
\vert a_2q_2\rangle
\label{3.62}\\
&{L_{coll}}_{a_1q_1a_2q_2a_3q_3a_4q_4} \stackrel{(3.53)}{=}
\delta_{a_2a_4} \delta_{q_2q_4} \delta_{a_1a_3} \delta_{q_1q_3}
\left( \langle \langle {\cal E}(q_1) \rangle\rangle
- \langle \langle {\cal E}(q_2) \rangle\rangle \rule{0mm}{4mm}\right)
\label{3.62'}\tag{3.62'}
\end{align}
Using the matrix-representations (\ref{3.16}), (\ref{3.62'}), one easily
checks that the following relations hold:
\begin{align}
&[\hat{P}, \hat{L}_{coll}] = 0 =  [1-\hat{Q}, \hat{L}_{coll}]
=-[\hat{Q}, \hat{L}_{coll}]
\label{3.63}
\end{align}
Due to the relations (\ref{3.63}), the NZ-equation (\ref{3.39}) can be
written in the form:
\begin{align}
\frac{d\hat{\rho}_P(t)}{dt}=&
 -\frac{i}{\hbar} \hat{P}(t)
 \left(\hat{L}_{cpl}(t)\hat{G}(t,0) \hat{\rho}_Q(0)
 +\left(\hat{L}_{coll}(t) +\hat{L}_{cpl}(t)\right) \hat{\rho}_P(t) \right)
-\frac{1}{\hbar^2} \hat{P}(t)\hat{L}_{cpl}(t)
 \int_0^{\infty} ds \,\hat{G}(t,s)\hat{Q}(s) \hat{L}_{cpl}(s) \hat{\rho}_P(s)
\notag\\
 &+\frac{i}{\hbar} \hat{P}(t)\hat{L}_{cpl}(t)
 \int_0^{\infty} ds \,\hat{G}(t,s) \dot{\hat{P}}(s)\hat{\rho}_P(s)
+\dot{\hat{P}}(t)\hat{\rho}_P(t) \label{3.64}
\end{align}
Eq. (\ref{3.64}) is still rigorously valid. We shall now apply the
``\underline{Markov approximation}".

This approximation is based on the observation that the non-negligible
contributions to the integrals in Eq. (\ref{3.64}) originate from a short
time interval $t-\tau_{rel} < s \leq t$, where the ``relaxation time"
$\tau_{rel}$ is much smaller than the typical time scale $\tau_c$ of the
collective motion.\\
One may estimate $\tau_{rel}$ to be between $10^{-23}$ s and $10^{-22}$ s and
$\tau_c$ to be of the order of $10^{-21}$ s, i.e.
\begin{equation}\label{3.65}
\tau_{rel} \ll \tau_c
\end{equation}
The quantity $\hat{\rho}_P(t)$ and the temperature $T(q,t)$
(see Section \ref{determination}) vary slowly, i.e.
on the time scale $\tau_c$, and so do the projection operators
$\hat{P}(t)$, $\hat{Q}(t)$, and the Liouvillean $\hat{L}_{cpl}(t)$, which
depend on time through the temperature.\\
The Markov approximation consists of several steps:
\begin{enumerate}
\item The first step is to replace the time variable $s$ by $t$ in the slowly
time-dependent quantities of the integrals:
\begin{align}
& \int_0^{\infty} ds \,\hat{G}(t,s) \hat{Q}(s) \hat{L}_{cpl}(s)
\hat{\rho}_P(s) \simeq
\int_0^{\infty} ds \,\hat{G}(t,s) \hat{Q}(t) \hat{L}_{cpl}(t)
\hat{\rho}_P(t)
\label{3.66} \\
& \int_0^{\infty} ds \,\hat{G}(t,s) \dot{\hat{P}}(s)\hat{\rho}_P(s) \simeq
\int_0^{\infty} ds \,\hat{G}(t,s) \dot{\hat{P}}(t)\hat{\rho}_P(t)
\label{3.67}
\end{align}
\item \underline{Simplification of the Green propagator $\hat{G}(t,s)$}\\
Replacing the slowly time-dependent quantities $\hat{Q}(\tau)$  and
$\dot{\hat{P}}(\tau)$ by  $\hat{Q}(t)$  and $\dot{\hat{P}}(t)$ in
(\ref{3.40}), the time-ordering $\hat{T}$ becomes superfluous and we obtain
\begin{equation}\label{3.68}
\hat{G}(t,s)\simeq \theta_0(t-s)  \hat{Q}(t) e^{ \ds
-\frac{i}{\hbar}(t-s)\left[\hat{L}\hat{Q}(t)-
\frac{\hbar}{i}\dot{\hat{P}}(t)\right]}
\end{equation}
As the operator $\dot{\hat{P}}(t)$ is proportional to $\dot{T}$ and as the
temperature varies slowly as a function of time, the change of the
temperature during the relaxation time, i.e. $\tau_{rel}\cdot \dot{T}(t)$ is
a very small quantity. We, therefore, neglect the term proportional to
$\dot{\hat{P}}(t)$ in (\ref{3.68}) altogether:
\begin{equation}\label{3.69}
\hat{G}(t,s)\simeq \theta_0(t-s)  \hat{Q}(t) e^{ \ds
-\frac{i}{\hbar}(t-s)\hat{L}\hat{Q}(t)}
\end{equation}
The form (\ref{3.69}) of the propagator is still too complicated for a
practical application because it contains the Liouvillean $\hat{L}$ of the
total system.\\
There are two possibilities for further simplification: Either one considers
the coupling Hamiltonian $\hat{H}_{cpl}$ as a perturbation, or one
approximates the propagator (\ref{3.69}) by considering the matrix elements
of $\hat{H}_{cpl}$ as random numbers. The two ways of proceeding are in a
certain sense complementary because the random matrix method implies that the
interaction $\hat{H}_{cpl}$ acts many times.\\
In what follows, we shall use the perturbation theory for $\hat{H}_{cpl}$. In
fact, perturbation theory  underlies most of the work on transport processes
and especially the work on nuclear fission \cite{Hofmann,Swiatecki-Blocki}.
The random matrix method was used by Weidenm\"uller et al. \cite{Weidenmuller}
for formulating a transport theory of heavy ions reactions.
It would be very interesting indeed to investigate the random matrix
approximation of the propagator (\ref{3.69}) parallely to the perturbation
treatment. The formulation of the propagator (\ref{3.69}) on the basis of the
random matrix theory and the consequences thereof, for instance concerning
the fluctuation-dissipation relation, would constitute a problem of its own
merit, and exceed the scope of the present work.

The perturbation theory implies that we replace the total Liouvillean
$\hat{L}$ in Eq. (\ref{3.69}) by the Liouvillean $\hat{L}_1$ corresponding to
the Hamiltonian $\hat{H}_1$ introduced in Eq. (\ref{3.50}). One thus replaces
the Green fuction (\ref{3.69}) by
\begin{equation}\label{3.70}
\hat{G}(t,s)\simeq \theta_0(t-s)  \hat{Q}(t) e^{ \ds
-\frac{i}{\hbar}(t-s)\hat{L}_1\hat{Q}(t)}
\end{equation}
As the basis functions $\vert aq\rangle$ are eigenfunctions of the
Hamiltonian $\hat{H}_1$ (see Eq. (\ref{3.51})), the propagator (\ref{3.70})
becomes very simple if we may replace the projection operator $\hat{Q}(t)$ in
the exponent of (\ref{3.70}) by 1:
\begin{equation}\label{3.71}
\hat{G}(t,s)\simeq \theta_0(t-s)  \hat{Q}(t)
e^{ \ds -\frac{i}{\hbar}(t-s)\hat{L}_1}
\end{equation}
The replacement of Eq. (\ref{3.70}) by Eq. (\ref{3.71}) would be strictly
valid if the projection operators $\hat{P}(t)$ and $\hat{Q}(t)$ commuted with
$\hat{L}_1$. One easily checks the result
\begin{align}
\left(\left[\hat{P},\hat{L}_1\right]\right)_{a_1q_1a_2q_2a_3q_3a_4q_4} =&
\delta_{a_1a_2} \delta_{a_3a_4} \delta_{q_1q_3} \delta_{q_2q_4} F_{a_2}(q_2)
\left({\cal E}_{a_1}(q_2)-{\cal E}_{a_1}(q_1)
+{\cal E}_{a_3}(q_1)-{\cal E}_{a_3}(q_2)\rule{0mm}{4mm}\right)
\label{3.72}
\end{align}
It implies that the matrix element
$\langle a_1q_1\vert \left[\hat{P},\hat{L}_1\right] \vert a_2q_2\rangle$,
and consequently also
$\langle a_1q_1\vert \left[\hat{Q},\hat{L}_1\right] \vert a_2q_2\rangle$
vanish for $q_1$=$q_2$.
As the propagator $\hat{G}(t,s)$ appears in the ``dissipative" integrals of
Eq. (\ref{3.64}), the contributions of the matrix-elements
$\langle a_1q_1\vert\hat{G}(t,s)\vert a_2q_2\rangle$  with $q_1$=$q_2$ are
expected to be much more important than the ones with $q_1$$\neq$$q_2$. As,
within the subspace of matrix elements diagonal in the g.c., the Liouvillean
$\hat{L}_1$ commutes with the projection operators, the approximation
(\ref{3.71}) of the expression (\ref{3.70}) is acceptable.
\end{enumerate}

Introducing the approximations (\ref{3.66}), (\ref{3.67}), and (\ref{3.71})
into Eq. (\ref{3.64}), the NZ-equation assumes the form :
\begin{align}
\frac{d\hat{\rho}_P(t)}{dt}&=
 -\frac{i}{\hbar} \hat{P}(t)
 \left[\hat{L}_{cpl}(t)e^{ \ds -\frac{i}{\hbar}t \hat{L}_1} \hat{\rho}_Q(0)
+\left(\hat{L}_{coll}(t) +\hat{L}_{cpl}(t)\right) \hat{\rho}_P(t)
\right]+\dot{\hat{P}}(t)\hat{\rho}_P(t)
\notag\\&
+\hat{P}(t)\hat{L}_{cpl}(t)
\left[ -\frac{1}{\hbar^2} \int_0^{\infty} ds  \theta_0(t-s)\hat{Q}(t)
e^{ \ds -\frac{i}{\hbar}(t-s)\hat{L}_1} \hat{L}_{cpl}(t) \hat{\rho}_P(t)
\right. \notag\\ & \hspace{20mm}\left.
 +\frac{i}{\hbar}  \int_0^{\infty} ds \,\theta_0(t-s)
\hat{Q}(t) e^{ \ds -\frac{i}{\hbar}(t-s) \hat{L}_1}
 \dot{\hat{P}}(t)\hat{\rho}_P(t)\right]
 \label{3.73}
\end{align}
Henceforth, we shall omit the inhomogeneous term in (\ref{3.73}).
Furthermore, in accordance with the approximation (\ref{3.71}) of the Green
function, we have put
\begin{equation}\label{3.74}
\hat{Q}(t)e^{ \ds -\frac{i}{\hbar}(t-s)\hat{L}_1} \hat{Q}(t)\simeq
\hat{Q}(t)e^{ \ds -\frac{i}{\hbar}(t-s)\hat{L}_1}
\end{equation}
in the dissipative term of Eq. (\ref{3.73}).\\
In the last term of Eq. (\ref{3.73}) we may leave away the projection
operator $\hat{Q}(t)$ making use of the identity (\ref{3.35}).
We can interpret this last term of (\ref{3.35}) as a modification of the
dissipative term due to the time-dependence of the temperature.
As the temperature changes slowly as a function of time, we expect this term
to be small.

We introduce the interaction representation of a Schr{\"o}dinger operator
$\hat{O}(t)$ by
\begin{equation}\label{3.75}
\hat{O}^{I}(s-t;t)=  e^{ \ds \frac{i}{\hbar}(s-t)\hat{L}_1} \hat{O}(t)
=e^{ \ds \frac{i}{\hbar}(s-t)\hat{H}_1}    \hat{O}(t)
e^{ \ds -\frac{i}{\hbar}(s-t)\hat{H}_1}
\end{equation}
The explicit time-dependence on $t$ of the projection operators and of the
Hamiltonian operators $\hat{H}_{coll}$ and $\hat{H}_{cpl}$ are produced by
the variation in time of the temperature. Whereas the dependence of
$\hat{O}^{I}(s-t;t)$ on the second argument $t$ is slow, the dependence on
$(s-t)$ is rapid, because it is related to the intrinsic excitations of the
system. It is satisfactory that the two different time scales are seen to
enter the equation of motion in a natural and automatic way:
\begin{align*}
\frac{d\hat{\rho}_P(t)}{dt}&=
 -\frac{i}{\hbar} \hat{P}(t)
 \left[\hat{H}_{coll}(t) + \hat{H}_{cpl}(t), \hat{\rho}_P(t) \right]
  +  \dot{\hat{P}}(t)\hat{\rho}_P(t)    \\
 & -\frac{1}{\hbar^2} \hat{P}(t)
 \int_0^{\infty} ds \, \theta_0(t-s)
\left[ \hat{H}_{cpl}(t),  \hat{Q}(t)
\left[ \hat{H}^{I}_{cpl}(s-t;t), \hat{\rho}^{I}_P(s-t;t) \right]\right]\\
 &+\frac{i}{\hbar}   \hat{P}(t) \int_0^{\infty}\!\!\!\! ds \,\theta_0(t-s)
 \left[ \hat{H}_{cpl}(t),  \hat{Q}(t)
  e^{i(s-t)\hat{H}_1/\hbar}  \dot{\hat{P}}(t)\hat{\rho}_P(t)
  e^{ -i(s-t)\hat{H}_1/\hbar}  \right]
\end{align*}
Due to Eq. (\ref{3.34}) and the relations $\hat{P}(t)\hat{Q}(t)=0$ and
$[\hat{Q}(t), \hat{L}_1]=0$, we can simplify the two last terms of this
equation. The NZ-equation for the density operator $\hat{\rho}_P(t)$ thus
reads:
\begin{align}
\frac{d\hat{\rho}_P(t)}{dt}=&
 -\frac{i}{\hbar} \hat{P}(t)
 \left[\hat{H}_{coll}(t) + \hat{H}_{cpl}(t), \hat{\rho}_P(t) \right]
  +  \dot{\hat{P}}(t)\hat{\rho}_P(t)   \notag\\
 & -\frac{1}{\hbar^2} \hat{P}(t)
 \int_0^{\infty} ds \, \theta_0(t-s)
\hat{H}_{cpl}(t)  \hat{Q}(t)
\left[ \hat{H}^{I}_{cpl}(s-t;t), \hat{\rho}^{I}_P(s-t;t) \right]\notag\\
 &+\frac{i}{\hbar}   \hat{P}(t) \int_0^{\infty} ds \,\theta_0(t-s)
\hat{H}_{cpl}(t)  e^{i(s-t)\hat{H}_1/\hbar}  \dot{\hat{P}}(t)\hat{\rho}_P(t)
  e^{ -i(s-t)\hat{H}_1/\hbar}
\label{3.76}
\end{align}
It is straightforward to derive from Eq. (\ref{3.76}) an equation of
motion for the reduced density matrix $R(q_1,q_2;t)$ defined by Eq.
(\ref{3.3}): Writing (\ref{3.76}) in terms of
a matrix representation and using (\ref{3.42}), we obtain
\begin{align*}
 \delta_{a_1a_2} F_{a_2}(q_2,T(q_2,t)) \sum_{a}
\langle a q_1\vert \frac{d\hat{\rho}(t)}{dt} \vert a q_2\rangle=&
\delta_{a_1a_2} F_{a_2}(q_2,T(q_2,t)) \left\{
-\frac{i}{\hbar} \sum_{a} \langle a q_1\vert
 \left[\hat{H}_{coll}(t) + \hat{H}_{cpl}(t), \hat{\rho}_P(t) \right]
 \vert a q_2\rangle   \right.\\
 & \left.-\frac{1}{\hbar^2}
 \int_0^{\infty} ds \, \theta_0(t\!-\!s) \sum_{a}  \langle a q_1\vert
\hat{H}_{cpl}(t)  \hat{Q}(t)
\left[ \hat{H}^{I}_{cpl}(s\!-\!t;t), \hat{\rho}^{I}_P(s\!-\!t;t) \right]
\vert a q_2\rangle \right.\\
 &\left.+\frac{i}{\hbar}  \int_0^{\infty} ds \,\theta_0(t\!-\!s)
\sum_{a} \langle a q_1\vert \hat{H}_{cpl}(t)  e^{i(s-t)\hat{H}_1/\hbar}
\dot{\hat{P}}(t)\hat{\rho}_P(t) e^{ -i(s-t)\hat{H}_1/\hbar}
\vert a q_2\rangle      \right\}
\end{align*}
Summing over $a_1,a_2$  and using the relation
$\sum_{a}F_a(q,T(q,t))$=1,
we find the following equation of motion for the reduced density matrix:
\begin{align}
\frac{dR(q_1, q_2 ; t)}{dt}=&
-\frac{i}{\hbar} \sum_{a_3q_3}
 \left\{{\cal H}_{a_3a_3}(q_1,q_3) R(q_3,q_2;t) F_{a_3}(q_2,T(q_2,t))
 - R(q_1,q_3;t)  F_{a_3}(q_3,T(q_3,t)) {\cal H}_{a_3a_3}(q_3,q_2)
\rule{0mm}{4mm}\right\}
\notag\\&
-\frac{1}{\hbar^2}  \int_0^{\infty} ds \, \theta_0(t-s) \sum_{aa_3q_3q_4}
{{\cal H}_{cpl}}_{aa_3}(q_1q_3;t) \left\{
{{\cal H}^{I}_{cpl}}_{a_3a}(q_3q_4) R_{aa}^{I}(q_4q_2;s-t,t) F_{a}(q_2,T(q_2,t))
\rule{0mm}{4mm}\right. \notag\\ & \left.\rule{0mm}{4mm} \hspace{60mm}
 - R_{a_3a_3}^{I}(q_3q_4;s-t,t)
F_{a_3}(q_4,T(q_4,t)) {{\cal H}^{I}_{cpl}}_{a_3a}(q_4q_2)  \right\}
\notag\\&
+\frac{1}{\hbar^2}  \int_0^{\infty} ds \, \theta_0(t-s) \sum_{aa_4q_3q_4}
{{\cal H}_{cpl}}_{aa}(q_1q_3;t) F_{a}(q_2,T(q_2,t))
\notag\\ & \hspace{-12mm} \times
\left\{ {{\cal H}^{I}_{cpl}}_{a_4a_4}(q_3q_4)
R_{a_4a_4}^{I}(q_4q_2;s-t,t) F_{a_4}(q_2,T(q_2,t))
- R_{a_4a_4}^{I}(q_3q_4;s-t,t)
 F_{a_4}(q_4,T(q_4,t)) {{\cal H}^{I}_{cpl}}_{a_4a_4}(q_4q_2)  \right\}
\notag\\&
 +\frac{i}{\hbar}  \int_0^{\infty} ds \,\theta_0(t-s)
\sum_{a_3q_3} {{\cal H}_{cpl}}_{a_3a_3}(q_1q_3) R_{a_3a_3}^{I}(q_3q_2;s-t,t)
    \dot{F}_{a_3}(q_2,T(q_2,t))
\label{3.77}
\end{align}
Here, we have introduced the following notations:
\begin{align}
&{\cal H}_{a_1a_2}(q_1,q_2) := \langle a_1q_1\vert \hat{H}\vert a_2q_2\rangle
\label{3.78} \\
&{{\cal H}_{cpl}}_{a_1a_2}(q_1q_2;t) := \langle a_1q_1\vert
\hat{H}_{cpl} (t) \vert a_2q_2\rangle \label{3.79} \\
&{{\cal H}^{I}_{cpl}}_{a_1a_2}(q_1q_2) := \langle a_1q_1\vert
\hat{H}^{I}_{cpl} (s-t;t) \vert a_2q_2\rangle
=e^{i(s-t)\omega_{a_1a_2}(q_1,q_2)} \langle a_1q_1\vert
\hat{H}_{cpl} (t) \vert a_2q_2\rangle
\label{3.80} \\
&R_{a_1a_2}^{I}(q_1q_2;s-t,t) :=
\langle a_1q_1\vert \hat{R}^{I}(s-t;t) \vert a_2q_2\rangle
= \delta_{a_1a_2} e^{i(s-t)\omega_{a_1a_1}(q_1,q_2)}
R(q_1,q_2;t) \label{3.81} \\
&\hat{R}(t):= \sum_{a'q'_1q'_2}\vert a'q'_1\rangle R(q'_1,q'_2;t)
\langle a'q'_2\vert \label{3.82} \\
&\langle a_1q_1\vert \hat{R}(t)\vert a_2q_2\rangle =\delta_{a_1a_2}
R(q_1,q_2;t) \label{3.83}
\end{align}
The frequency $\omega_{a_1a_2}(q_1,q_2)$ occurring in Eq. (\ref{3.81}) is
related to the difference of eigen-energies (\ref{3.52}) by
\begin{equation}\label{3.81'}\tag{3.81'}
\hbar\omega_{a_1a_2}(q_1,q_2):={\cal E}_{a_1}(q_1)- {\cal E}_{a_2}(q_2)
\end{equation}
 In the last term of Eq.(\ref{3.77}), the factor $\dot{F}_{a_3}$ can
be written in the form
\begin{equation}\label{3.84}
 \dot{F}_{a_3}(q_2,T(q_2,t))=\frac{\partial F_{a_3}(q_2,T(q_2,t))}
 {\partial T} \dot{T}(q_2,t)
\end{equation}
The equation of motion (\ref{3.77}) for the reduced density matrix
$R(q_1,q_2;t)$ must be solved together with the equation defining $T(q,t)$
which will be derived in the next Section.

\section{Determination of the temperature}  \label{determination}
\setcounter{equation}{0}

So far, the temperature entered the ``ansatz" for $\hat{\rho}_P(t)$ as an
undetermined parameter. If we had exact solutions for $\hat{\rho}_P(t)$ and
$\hat{\rho}_Q(t)$, the choice of the temperature would not matter, as it
would only concern the decomposition of $\hat{\rho}(t)$ into
$\hat{\rho}_P(t)$ and $\hat{\rho}_Q(t)$. Of course, in reality, we determine
$\hat{\rho}_P(t)$ from the equation (\ref{3.76}), which was derived using the
Markov approximation. One of the prerogatives of this approximation is that
$\hat{\rho}_P(t)$ should depend slowly on time. Even more so, we require
physically that $\hat{\rho}_P(t)$ should represent the slowly time-dependent
part of the density operator $\hat{\rho}(t)$ ``as accurately as possible". It
is this requirement which must determine the choice of the temperature.

As the mean total energy of the system is independent of time, we should
postulate in any case that the total energy of the system
\begin{equation}\label{4.1}
E := {\rm Tr}\left(\hat{H} \hat{\rho}(t)\right)
={\rm Tr} \left(\hat{H} \hat{\rho}_P^h(t)\right)
\end{equation}
is independent of time,
where
\begin{equation}\label{4.2}
\hat{\rho}_P^h(t)=\frac{1}{2}\left(\hat{\rho}_P(t)+\hat{\rho}_P^\dagger (t)\right)
\end{equation}
is the hermitian part of $\hat{\rho}_P$.\\
The condition (\ref{4.1}) was used in Ref.~\cite{Hofmann} to determine the
temperature as a function of time.
In addition, at any given time $t$, the temperature $T$ should also depend on
the g.c. $q$, because the amount of intrinsic excitation is expected to vary
along the dynamical paths of the system.
In particular, we expect it to rise as the system approaches the scission
region. Therefore, as a natural further condition to be fulfilled by the
hermitian part $\hat{\rho}_P^h(t)$ of the reference density
$\hat{\rho}_P(t)$, we require that it should yield
\underline{the same average energy as the total density
$\hat{\rho}(t)$ at any given value of the g.c. $q$}.

In order that this condition involves only \underline{real} functions, it is
given in the form
\begin{align}
&\frac{1}{2}\sum_{a} {\rm Re}\left\{
\langle aq\vert \left[\hat{H},\hat{\rho}(t)\right]_{+}\vert aq\rangle
\right\}
=\frac{1}{2}\sum_{a} {\rm Re}\left\{
\langle aq\vert \left[\hat{H},\hat{\rho}_P^h(t)\right]_{+}\vert aq\rangle
\right\}
\label{4.3}
\end{align}
The symmetrized forms $\dfrac{1}{2} \left[\hat{H},\hat{\rho}\right]_{+}$ and
$\dfrac{1}{2}\left[\hat{H},\hat{\rho}_P^h\right]_{+}$ are used in
(\ref{4.3}) because the simple products $\hat{H}\hat{\rho}$ and
$\hat{H}\hat{\rho}_P^h$ are not
hermitian. In addition, we choose the real parts of the matrix elements
because, in case of the biorthogonal basis,
$\langle \widetilde{\phi_a(q)}\vert \left[\hat{H},\hat{\rho}(t)\right]_{+}
\vert \phi_a(q)\rangle$  and
$\langle \widetilde{\phi_a(q)}\vert
\left[\hat{H},\hat{\rho}_P^h(t)\right]_{+} $ $ \vert \phi_a(q)\rangle$  are
not necessarily real.  \\
Summing the Eq. (\ref{4.3}) over the g.c. $q$ and using the completeness
relation $\ds\sum_{aq}\vert aq\rangle\langle aq\vert $=1, one finds that
the Eq. (\ref{4.3}) yields the condition (\ref{4.1}) of a conserved mean
value of the total energy.

The condition (\ref{4.3}) can also be written in the form
\begin{equation}\label{4.4}
\sum_{a} {\rm Re}\left\{
\langle aq\vert \left[\hat{H},\hat{\rho}_Q^h(t)\right]_{+}\vert aq\rangle
\right\}=0
\end{equation}
where $\hat{\rho}_Q^h(t)$ is defined by
\begin{equation}\label{4.5}
\hat{\rho}_Q^h(t) := \hat{\rho}(t) - \hat{\rho}_P^h(t)
\end{equation}
For the sake of simplicity, in Eq. (\ref{4.4}), we replace the hermitian part
$\hat{\rho}_Q^h(t)$ by $\hat{\rho}_Q(t)$ :
\begin{equation}\label{4.6}
\sum_{a} {\rm Re}\left\{
\langle aq\vert \left[\hat{H},\hat{\rho}_Q(t)\right]_{+}\vert aq\rangle
\right\}=0
\end{equation}
If we use the orthonormal basis functions
($\vert aq\rangle = \vert \Psi_a(q)\rangle$,
$\langle aq\vert  = \langle  \Psi_a(q)\vert$), the conditions (\ref{4.4}) and
(\ref{4.6}) are exactly equivalent, as one can show that
\begin{equation}\label{4.7}
\sum_{a} {\rm Re}\left\{
\langle  \Psi_a(q) \vert \left[\hat{H},\hat{\rho}_Q^a(t)\right]_{+}
\vert \Psi_a(q) \rangle \right\}=0
\end{equation}
where
\begin{equation}\label{4.8}
\hat{\rho}_Q^a(t)= \frac{1}{2}\left(\hat{\rho}_Q(t) -
\hat{\rho}_Q^\dagger (t)\right)
\end{equation}
is the antihermitian part of $\hat{\rho}_Q(t)$.\\
If we use the biorthogonal basis functions
($\vert aq\rangle = \vert \phi_a(q)\rangle$,
$\langle aq\vert  = \langle  \widetilde{\phi_a(q)}\vert$), the condition
(\ref{4.6}) differs from (\ref{4.4}), because
\begin{equation}\label{4.9}
\sum_{a} {\rm Re}\left\{
\langle \widetilde{\phi_a(q)} \vert \left[\hat{H},\hat{\rho}_Q^a(t)\right]_{+}
\vert \phi_a(q) \rangle \right\} \neq 0
\end{equation}
If the spacing $\Delta q$ is such that the overlap parameter $\varepsilon
\ll 1$, the expression (\ref{4.9}), which is of ${\cal O}(\varepsilon)$, is
small. We thus expect that the simpler condition (\ref{4.6}) yields almost
the same temperature $T(q,t)$ as condition (\ref{4.4}).
Therefore, we, henceforth, use the relation (\ref{4.6}) for determining the
function $T(q,t)$.

Let us note that summing condition (\ref{4.6}) over $q$ leads to the
conservation of the total energy
\begin{equation}\label{4.10}
{\rm Re}\left\{{\rm Tr}\left(\hat{H} \hat{\rho}(t)\right) \right\}
= {\rm Re}\left\{{\rm Tr}\left(\hat{H} \hat{\rho}_P(t)\right) \right\}
= {\rm Re}\left\{{\rm Tr}\left(\hat{H} \hat{\rho}_P^h(t)\right) \right\}
\end{equation}
as we have
\begin{equation}\label{4.11}
{\rm Re}\left\{{\rm Tr}\left(\hat{H} \hat{\rho}_P^{a}(t)\right) \right\}
= \frac{1}{2}\left\{{\rm Tr}\left(\hat{H} \hat{\rho}_P^a(t)\right)
+ {\rm Tr}\left(\hat{H} \hat{\rho}_P^a(t)\right)^\dagger \right\} =0
\end{equation}
Let us now investigate the condition (\ref{4.6}) in more detail:\\
The equation (\ref{3.37}) presents the density operator $\hat{\rho}_Q(t)$ as
a formal solution of the equation of motion (\ref{3.25}).
Applying the Markov approximation (\ref{3.71}) to the Green operator
$\hat{G}(t,s)$, leaving away the memory term
$\hat{G}(t,0)\hat{\rho}_Q(0)$, and replacing the source term
$\hat{\sigma}(s)$ in Eq. (\ref{3.37}) by $\hat{\sigma}(t)$, we obtain the
density operator  $\hat{\rho}_Q(t)$ in the form
\begin{equation}\label{4.12}
\hat{\rho}_Q(t) \simeq  \int_0^{\infty} ds \, \theta_0(t-s)  \hat{Q}(t)
 e^{ \ds -\frac{i}{\hbar}(t-s)\hat{L}_1} \hat{\sigma}(t)
\end{equation}
As for the condition (\ref{4.6})
\begin{align}
 {\rm Re} \left\{ \sum_{a}
\langle aq\vert \left[\hat{H},\hat{\rho}_Q(t)\right]_{+}\vert aq\rangle
\right\}=&
 {\rm Re} \left\{ \sum_{aa_1q_1}   \left(
\langle aq \vert \hat{H} \vert a_1q_1 \rangle
\langle a_1q_1 \vert \hat{\rho}_Q(t) \vert aq \rangle
\right.\right.\notag \\ &\left. \left.
+ \langle aq \vert \hat{\rho}_Q(t) \vert a_1q_1 \rangle
\langle a_1q_1 \vert \hat{H} \vert aq \rangle
\right)\right\}=0
\label{4.13}
\end{align}
it is convenient to decompose  $\hat{\rho}_Q(t)$ into two terms of different
physical origin which correspond to the two terms of the source term
(\ref{3.31}):
\begin{align}
& \hat{\rho}_Q(t) =\hat{\rho}_Q^{(1)}(t) +\hat{\rho}_Q^{(2)}(t)
\label{4.14}    \\
& \hat{\rho}_Q^{(1)}(t) :=-\frac{i}{\hbar}
 \int_0^{\infty} ds \, \theta_0(t-s)  \hat{Q}(t)
 e^{ \ds -\frac{i}{\hbar}(t-s)\hat{L}_1}
\hat{Q}(t)\hat{L}_{cpl}(t)\hat{\rho}_P(t)
\label{4.15}    \\
& \hat{\rho}_Q^{(1)}(t) \simeq -\frac{i}{\hbar}
 \int_0^{\infty} ds \, \theta_0(t-s)  \hat{Q}(t)
 e^{ \ds -\frac{i}{\hbar}(t-s)\hat{L}_1} \hat{L}_{cpl}(t)\hat{\rho}_P(t)
\label{4.16}    \\
& \hat{\rho}_Q^{(2)}(t) := -
 \int_0^{\infty} ds \, \theta_0(t-s)
 e^{ \ds -\frac{i}{\hbar}(t-s)\hat{L}_1}   \dot{\hat{P}}(t)\hat{\rho}_P(t)
\label{4.17}
\end{align}
Replacing the form (\ref{4.15}) of $\hat{\rho}_Q^{(1)}(t)$ by the form
(\ref{4.16}) implies the same approximation which was used to replace the
form (\ref{3.70}) of the Green operator by the simpler form (\ref{3.71}).
It is justified because, due to the factor
$\langle aq \vert \hat{H} \vert a_1q_1 \rangle$,
the main contribution to the condition (\ref{4.13}) arises for $q_1$=$q$.
In the subspace of equal g.c., the projection operators commute with the
Liouvillean $\hat{L}_1$, as one can see from Eq. (\ref{3.72}). \\
After a straightforward calculation, one obtains
\begin{align}
 \langle a_1q_1\vert \hat{\rho}_Q^{(1)}(t)\vert aq\rangle
=\!&-\frac{i}{\hbar} \! \int_0^{\infty} \!\!\!ds \, \theta_0(t\!-\!s)
\left\{  \mbox{\rule[-2mm]{0mm}{9mm}}
\langle a_1q_1\vert \left[ \hat{H}_{cpl}^{I}(s\!-\!t,t),
\hat{\rho}_P^{I} (s\!-\!t,t) \right] \vert aq\rangle
\right. \notag\\ & \left.\hspace{-22mm}
-\delta_{a_1a} F_a(q,T(q,t)) \sum_{b}
\langle bq_1\vert \left[ \hat{H}_{cpl}^{I}(s-t,t),
\hat{\rho}_P^{I} (s-t,t) \right] \vert bq\rangle \right\}
\label{4.18} \\
 \langle a_1q_1\vert \hat{\rho}_Q^{(2)}(t)\vert aq\rangle
=&- \!\!\int_0^{\infty}\!\!ds \, \theta_0(t-s) R^{I}_{a_1a}(q_1,q;s-t,t)
\frac{\partial F_a(q,T(q,t))}{\partial T} \dot{T}(q,t)
\label{4.19}
\end{align}
where $\hat{H}_{cpl}^{I}(s-t,t)$ and $\hat{\rho}_P^{I} (s-t,t)$ are the
operators $\hat{H}_{cpl}$ and $\hat{\rho}_P$ in the interaction picture (see
Eq. (\ref{3.75})), and where $R^{I}_{a_1a}(q_1,q;s-t,t)$  is defined by Eq.
(\ref{3.81}).  \\
The temperature occurs in the condition (\ref{4.13}) with different generator
coordinates as arguments. As the matrices of $\hat{\rho}_Q^{(1)}$ and
$\hat{\rho}_Q^{(2)}$ are multiplied with matrix-elements
${\cal H}_{aa_1}(q,q_1)$ and ${\cal H}_{a_1a}(q_1,q)$ which exhibit a narrow
Gaussian-like dependence on $q-q_1$, and as the ``local" temperature
$T(q_1,t)$ varies smoothly as a function of $q_1$, we may introduce the
approximation  of replacing all the temperature functions appearing in Eq.
(\ref{4.13}) by $T(q,t)$.\\
Let us introduce the matrices ${\cal T}^{(1)}$ and ${\cal T}^{(2)}$ defined
by putting all the $T$-functions occurring in the matrices
$\hat{\rho}_Q^{(1)}$ and $\hat{\rho}_Q^{(2)}$ by $T(q,t)$
\begin{align}
&{\cal T}^{(1)}_{a_1a}(q_1,q;t):= \left\{\langle a_1q_1\vert
\hat{\rho}_Q^{(1)}\vert aq\rangle \right\}_{{\mbox all\hspace{1mm}}T=T(q,t)}
\label{4.20} \\
&{\cal T}^{(2)}_{a_1a}(q_1,q;t)\dot{T}(q,t):= \left\{ \langle a_1q_1\vert
\hat{\rho}_Q^{(2)} \vert aq\rangle \right\}_{{\mbox all\hspace{1mm}}T=T(q,t)}
\label{4.21}
\end{align}
and correspondingly for the matrix-elements ${\cal T}^{(1)}_{aa_1}(q,q_1;t)$
and ${\cal T}^{(2)}_{aa_1}(q,q_1;t)$. As can be seen from Eq. (\ref{4.19}),
the r.h.s. of (\ref{4.21}) reads more explicitely
\begin{align}
&{\cal T}^{(2)}_{a_1a}(q_1,q;t)= \left\{
-\int_0^{\infty}\!\!\!ds \, \theta_0(t-s) R^{I}_{a_1a}(q_1,q;s-t,t)
\frac{\partial F_a(q,T)}{\partial T}
 \right\}_{T=T(q,t)}
\label{4.22}
\end{align}
The condition (\ref{4.6}) thus assumes the form
\begin{align}
& {\rm Re}   \sum_{a} \left(
{\left[ {\cal H}, {\cal T}^{(1)} \right]_{+}}_{aa}\!\!\!\!\!(q,q)
+ {\left[ {\cal H}, {\cal T}^{(2)} \right]_{+}}_{aa}\!\!\!\!\!(q,q)
\cdot \dot{T}(q,t) \right)=0 \label{4.23}
\end{align}
or, more explicitely,
\begin{align}
& {\rm Re}  \left\{ \sum_{aa_1q_1}   \left(
{\cal H}_{aa_1}(q,q_1) {\cal T}^{(1)}_{a_1a}(q_1,q;t) +
 {\cal T}^{(1)}_{aa_1}(q,q_1;t) {\cal H}_{a_1a}(q_1,q)\right)
\right.\notag \\ & \left.
    +  \sum_{aa_1q_1}   \left(
{\cal H}_{aa_1}(q,q_1) {\cal T}^{(2)}_{a_1a}(q_1,q;t) +
 {\cal T}^{(2)}_{aa_1}(q,q_1;t) {\cal H}_{a_1a}(q_1,q)\right)\dot{T}(q,t)
\right\}=0
\label{4.24}
\end{align}
The temperature condition (\ref{4.24}) and the equation of motion
(\ref{3.77}) represent a coupled system of equations for the reduced density
matrix $R(q_1,q_2;t)$ and the temperature $T(q,t)$. The solutions
$R(q_1,q_2;t)$ and $T(q,t)$ are uniquely determined whenever initial values
$R(q_1,q_2;t=0)$ and $T(q,t=0)$ are defined. As an example, one could assume
that the initial reference density $\hat{\rho}_P(0)$ should correspond to a
canonical distribution at the g.c. $q_0$ of the ground state valley.
Then the initial values would be
\begin{align}
& R(q_1,q_2;0) = \delta_{q_1q_0} \delta_{q_2q_0} \label{4.25} \\
&\langle a_1q_1\vert \hat{\rho}_P(0) \vert a_2q_2\rangle =
\delta_{q_1q_0} \delta_{q_2q_0} \delta_{a_1a_2} F_{a_1}(q_0,T(q_0,0))
 \label{4.26}
\end{align}
The initial temperature could be defined by the knowledge of the total energy
$E_{tot}$:
\begin{equation}\label{4.26k}
\sum_{a_1}\langle a_1q_0\vert \hat{H}\vert a_1q_0\rangle
F_{a_1}(q_0,T(q_0,0)) = E_{tot}
\end{equation}
Let us comment the numerical problem of evaluating solutions of the coupled
set of equations (\ref{3.77}) for the reduced density $R(q_1,q_2;t)$ and the
equation (\ref{4.24}) for the temperature $T(q,t)$:\\
Given the fact that a realistic description of the fission process requires a
set of 2 to 3 generator coordinates, the technical problem is one of storage
capacity or of rapid subroutines. As already mentioned, we replace the
continuous values of the g.c.s by a network of discrete values and we also
discretize the time. The step lengths $\Delta q$ and $\Delta t$ are to be
determined by the required accuracy. \\
From the knowledge of the initial functions $R(q_1,q_2;t=0)$ and $T(q,t=0)$,
we first obtain $R(q_1,q_2;t=\Delta t)$ from Eq. (\ref{3.77}) and then
$T(q,\Delta t)$ from Eq. (\ref{4.24}). Subsequently, at the
n$^{\mbox{th}}$ time step, one obtains $R(q_1,q_2;t_n+\Delta t)$ from
substituting $R(q_1,q_2;t\leq t_n)$, $T(q,t\leq t_n)$ on the r.h.s. of
Eq. (\ref{3.77}) and, analogously, one finds $T(q,t_n+\Delta t)$ from
Eq. (\ref{4.24}). If the storage capacity of the computer is sufficiently
great,
one can speed up the calculation considerably by storing the time-independent
matrices $\langle a_1q_1\vert \hat{H}_1\vert a_2q_2\rangle$,
$\langle a_1q_1\vert \hat{H}_{cpl}\vert a_2q_2\rangle$, and the
eigenvalues $\varepsilon_a(q)$ before entering the routines for solving the
Eqs. (\ref{3.77}) and (\ref{4.24}).\\
Through the Eqs. (\ref{3.77}) and (\ref{4.24}) the initial distribution
$R(q_1,q_2;t=0)$ is thus propagated through the potential landscape as a
function of time up to the scission configurations which represent a
$({\cal N}-1)$ surface in the space of ${\cal N}$ generator coordinates.
The propagation of the reduced density $R(q_1,q_2;0)$ up to values of
$R(q_1,q_2;t)$ on the ``scission surface" is reminiscent of a percolation
problem.

Let us conclude this Section with a few remarks concerning the entropy:\\
As a first step, we have to introduce an adequate definition of the entropy.
It seems to be appropriate to define a ``local" entropy $S(q,t)$ in
relation to the ``local" temperature $T(q,t)$:
\begin{align}
& S(q,T) := -\sum_{a} F_a(q,T) \,{\rm ln}  F_a(q,T) \label{4.27k} \\
& S(q,T) = \frac{\langle \langle {\cal E}'(q,T) \rangle \rangle -G(q,T)}{T}
\label{4.28k}
\end{align}
where
\begin{equation}\label{4.29k}
\langle \langle {\cal E}'(q,T) \rangle \rangle :=
\sum_{a} {\cal E}'_a(q) F_a(q,T)
\end{equation}
The total entropy S(t) of the system is then to be defined by summing
$S(q,T)$ over all nuclear shapes weighted with the probability to find the
system at the g.c. $q$ at time $t$:
\begin{equation}\label{4.30k}
S(t) := \sum_{q} S(q,T(q,t))\,R(q,q;t)
\end{equation}
We remind the reader of the fact (see Section 2) that $R(q,q;t)$ can only be
interpreted as a probability if we define it through the representation in
terms of the orthonormal basis states $\Psi_a(q)$.\\
As long as we neglect the emission of light particles and photons during the
fission process, we expect that the total entropy $S(t)$ rises as a function
of time
\begin{align}
&\frac{dS(t)}{dt} =\sum_{q} \frac{\partial S(q,T(q,t))}{\partial T}
\dot{T}(q,t) +\sum_{q} S(q,T(q,t))\,\dot{R}(q,q;t) \geq 0
\label{4.31k}
\end{align}
As the process of creating an increasing amount of intrinsic excitations,
which are simply described by a canonical distribution, corresponds to a loss
of information, the entropy must rise.
However, as the functions $T(q,t)$ and $R(q,q;t)$ in (\ref{4.31k}) can only
be obtained by solving the coupled equations (\ref{3.77}) and (\ref{4.24}),
we cannot prove the statement (\ref{4.31k}) in generality.\\
In a realistic description of the fission process, deexcitation processes
must be taken into account. As the emission processes reduce the entropy of
the remaining system, the statement (\ref{4.31k}) can no longer be made.

\section{The fluctuation-dissipation relation}  \label{fluctuation}  
\setcounter{equation}{0}

The fluctuation-dissipation theorem relates part of the Fourier transforms of
the response function and of the correlation function of a pair of operators
with each other \cite{fluc-diss,Hofmann,Grossmann}. The response and correlation
functions appear in a natural way when studying the response of a system in
thermal equilibrium with regard to small perturbations. As we chose to treat
the coupling Hamiltonian $\hat{H}_{cpl}$ in the integral term of the
NZ-equation (\ref{3.77}) as a perturbation, we may expect that some sort of
fluctuation-dissipation relation should hold. As we do not deal with a
thermal equilibrium, but with a slow passage of the system through thermal
equilibrium states pertaining to different values of the g.c., we expect that
a fluctuation-dissipation relation holds only in the limit of very slow
collective transport.\\
Apart from the last term in Eq. (\ref{3.77}), which depends on the time rate
$\dot{T}$ of the temperature, the time integral in the Eq. (\ref{3.77}) can
be written as a sum of two terms
\begin{align}
{\cal D}(q_1,q_2,t):=&\sum_{q'_1} D_1(q_1,q_2,q'_1,t)R(q'_1,q_2;t)+
\sum_{q'_1 q'_2} D_2(q_1,q_2,q'_1,q'_2,t)R(q'_1,q'_2;t)
  \label{5.1}
\end{align}
where $D_1(q_1,q_2,q'_1,t)$ and $D_2(q_1,q_2,q'_1,q'_2,t)$ are defined as
follows
\begin{align}
D_1(q_1,q_2&,q'_1,t):=-\frac{1}{\hbar^2} \sum_{aa'q'}
\int_0^{\infty} \!\!\!\!ds \,\,\, \theta_0(t-s) F_a(q_2,T(q_2,t))
e^{i(s-t)\omega_{aa}(q'_1,q_2)}  \notag\\
&\times\left\{ {{\cal H}_{cpl}}_{aa'}(q_1,q';t)
 {{\cal H}^{I}_{cpl}}_{a'a}(q',q'_1;s-t,t)
 -{{\cal H}_{cpl}}_{a'a'}(q_1,q';t)
 {{\cal H}^{I}_{cpl}}_{aa}(q',q'_1;s-t,t)
 F_{a'}(q_2,T(q_2,t))\right\}
  \label{5.2}\\
D_2(q_1,q_2&,q'_1,q'_2,t):=\frac{1}{\hbar^2} \sum_{aa'}
\int_0^{\infty} \!\!\!\!ds \,\,\, \theta_0(t-s) F_{a'}(q'_2,T(q'_2,t))
e^{i(s-t)\omega_{a'a'}(q'_1,q'_2)}  \notag\\
&\times\left\{ {{\cal H}_{cpl}}_{aa'}(q_1,q'_1;t)
      {{\cal H}^{I}_{cpl}}_{a'a}(q'_2,q_2;s-t,t)
 -{{\cal H}_{cpl}}_{aa}(q_1,q'_1;t)
      {{\cal H}^{I}_{cpl}}_{a'a'}(q'_2,q_2;s-t,t)
F_{a}(q_2,T(q_2,t))  \right\}
  \label{5.3}
\end{align}
The matrix-elements of the coupling Hamimtonian $\hat{H}_{cpl}$ depend on
the values of the generator coordinates in the bra and ket states through a
narrow Gaussian (see Eqs. (\ref{1.16}), (\ref{1.17})). Thus, the
matrix-elements have their largest values for equal values of the generator
coordinates in bra and ket. Let us, therefore, consider the parts
$D_1^{(0)}$ and $D_2^{(0)}$ of $D_1$ and $D_2$,
which are of zeroth order in the overlap parameter $\varepsilon$ (see Eq.
(\ref{2.6})):
\begin{align}
&D_1^{(0)}(q_1,q_2,q'_1,t)=-\frac{1}{\hbar^2} \sum_{aa'}
\int_0^{\infty} \!\!\!\!ds \,\,\, \theta_0(t-s) F_a(q_2,T(q_2,t))
e^{i(s-t)\omega_{aa}(q_1,q_2)}
\notag\\ & \hspace{2mm}\times
\left\{ {{\cal H}_{cpl}}_{aa'}(q_1,q_1;t)
 {{\cal H}^{I}_{cpl}}_{a'a}(q_1,q_1;s-t,t)
 -{{\cal H}_{cpl}}_{a'a'}(q_1,q_1;t)
 {{\cal H}^{I}_{cpl}}_{aa}(q_1,q_1;s-t,t)
 F_{a'}(q_2,T(q_2,t)) \right\}  \delta_{q_1q'_1}
  \label{5.4}\\
&D_2^{(0)}(q_1,q_2,q'_1,q'_2,t)=\frac{1}{\hbar^2} \sum_{aa'}
\int_0^{\infty} \!\!\!\!ds \,\,\, \theta_0(t-s) F_{a'}(q_2,T(q_2,t))
e^{i(s-t)\omega_{a'a'}(q_1,q_2)}
 \notag\\ & \hspace{2mm}\times
\left\{ {{\cal H}_{cpl}}_{aa'}(q_1,q_1;t)
        {{\cal H}^{I}_{cpl}}_{a'a}(q_2,q_2;s-t,t)
 -{{\cal H}_{cpl}}_{aa}(q_1,q_1;t)
         {{\cal H}^{I}_{cpl}}_{a'a'}(q_2,q_2;s-t,t)
  F_{a}(q_2,T(q_2,t)) \right\}
 \delta_{q'_1q_1}\delta_{q'_2q_2}
  \label{5.5}
\end{align}
The exponential factors in (\ref{5.2}) to (\ref{5.5})  (for their definition,
see Eqs. (\ref{3.81}), (\ref{3.81'})) refer to differences of energies at
\underline{different} $q$-values $(q_1,q_2)$ on the \underline{same}
potentiel energy surface characterized by given quantum numbers of intrinsic
excitation ``$a$'' or ``$a'$''.\\
In the special case $q_1$=$q_2$, the exponential factors are equal to 1 and the
coefficients assume the simple form
\begin{align}
&D_1^{(0)}(q_1,q_1,q'_1,t)=-\frac{1}{\hbar^2}
\int_0^{\infty} \!\!\!\!ds \,\,\, \theta_0(t-s)
\langle\langle\left({{\cal H}_{cpl}}(q_1,q_1)
  - \langle\langle {{\cal H}_{cpl}}(q_1,q_1) \rangle\rangle_{q_1}\right)
\notag\\ & \hspace{60mm}\cdot
\left( {{\cal H}^{I}_{cpl}}(q_1,q_1)
  - \langle\langle{{\cal H}^{I}_{cpl}}(q_1,q_1)
  \rangle\rangle_{q_1}\right)\rangle\rangle_{q_1} \delta_{q_1q'_1}
  \label{5.6}\\
&D_2^{(0)}(q_1,q_1,q'_1,q'_2,t)=\frac{1}{\hbar^2}
\int_0^{\infty} \!\!\!\!ds \,\,\, \theta_0(t-s)
\langle\langle\left({{\cal H}^{I}_{cpl}}(q_1,q_1)
  - \langle\langle {{\cal H}^{I}_{cpl}}(q_1,q_1) \rangle\rangle_{q_1}\right)
\notag\\ & \hspace{60mm}\cdot
\left( {{\cal H}_{cpl}}(q_1,q_1)
  - \langle\langle{{\cal H}_{cpl}}(q_1,q_1)
  \rangle\rangle_{q_1}\right)\rangle\rangle_{q_1}
 \delta_{q'_1q_1}\delta_{q'_2q_1}
  \label{5.7}
\end{align}
where ${{\cal H}_{cpl}}$ and ${{\cal H}^{I}_{cpl}}$ signify the matrices
\begin{align}
&{{\cal H}_{cpl}}(q_1,q_2):=
\left\{{{\cal H}_{cpl}}_{a_1a_2}(q_1,q_2;t)\right\}
  \label{5.8}\\
&{{\cal H}^{I}_{cpl}}(q_1,q_2):=
\left\{{{\cal H}^{I}_{cpl}}_{a_1a_2}(q_1,q_2;s-t,t)\right\}
  \label{5.9}
\end{align}
and where we denote the thermal mean-values of an operator $\hat{O}$ at a
given value of $q$ of the generator coordinates by
\begin{equation}\label{5.10}
 \langle\langle\hat{O}\rangle\rangle_{q} :=
 \sum_{a} F_a(q,T(q,t)) \langle aq\vert \hat{O}\vert aq\rangle
\end{equation}
The integrands on the right hand sides of the Eqs. (\ref{5.6}), (\ref{5.7})
are seen to represent the correlation between the fluctuations
$\delta H_{cpl}(t)$, $\delta H^{I}_{cpl}(s-t,t)$ of the operators
$\hat{H}_{cpl}(t)$, $\hat{H}^{I}_{cpl}(s-t,t)$ around their
respective thermal mean values.
Decomposing the product of the fluctuations into a commutator and an
anticommutator, we may write the coefficients (\ref{5.6}) and (\ref{5.7}) in
the form
\begin{align}
&D_1^{(0)}(q_1,q_1,q'_1,t)=-\frac{1}{2\hbar^2} \!\!
\int_0^{\infty} \!\!\!\!ds \, \theta_0(t-s)  \left\{
\langle\langle\left[\delta{\cal H}_{cpl}(t),
        \delta {\cal H}^{I}_{cpl}(s-t,t)\right]_-\rangle\rangle_{q}
+ \langle\langle\left[\delta{\cal H}_{cpl}(t),
        \delta {\cal H}^{I}_{cpl}(s-t,t)\right]_+\rangle\rangle_{q}
 \right\}\delta_{q_1q'_1}
  \label{5.11}\\
&D_2^{(0)}(q_1,q_1,q'_1,q'_2,t)=\frac{1}{2\hbar^2}  \!\!
\int_0^{\infty} \!\!\!\!ds \, \theta_0(t-s)  \left\{
\langle\langle\left[\delta {\cal H}^{I}_{cpl}(s-t,t),
        \delta{\cal H}_{cpl}(t)\right]_-\rangle\rangle_{q}
+ \langle\langle\left[ \delta {\cal H}^{I}_{cpl}(s-t,t),
       \delta{\cal H}_{cpl}(t)\right]_+\rangle\rangle_{q}
 \right\}  \delta_{q'_1q_1}\delta_{q'_2q_2}
  \label{5.12}
\end{align}
The mean-values of the commutator parts of Eqs. (\ref{5.11}) and (\ref{5.12})
are proportional to the ``\underline{response function}" defined by
\begin{align}
\chi(q,s';t):=&\frac{i}{\hbar} \sum_{a}  \theta_0(-s') F_a(q,T(q,t))
\langle aq\vert\left[\delta \hat{H}^{I}_{cpl}(s',t) ,
\delta \hat{H}_{cpl}(t)\right] \vert aq  \rangle \notag\\
=& \frac{i}{\hbar} \theta_0(-s')
\langle\langle\left[ \delta {\cal H}^{I}_{cpl}(s',t),
       \delta{\cal H}_{cpl}(t)\right]_-\rangle\rangle_{q}
  \label{5.13}
\end{align}
Here, we used the following definitions
\begin{align}
&s'=s-t   \label{5.14}\\
&\delta \hat{H}^{I}_{cpl}(s',t) := \hat{H}^{I}_{cpl}(s',t)-
\langle \langle \hat{H}^{I}_{cpl}(s',t) \rangle \rangle_q
  \label{5.15}  \\
& \delta \hat{H}_{cpl}(t):= \hat{H}_{cpl}(t)-
\langle \langle \hat{H}_{cpl}(t) \rangle \rangle_q
  \label{5.16}
\end{align}
The slow time-dependence on $t$ is produced by the dependence of the
temperature $T(q,t)$ on $t$.
The time-dependence on $s'$ is rapid due to the dependence on the intrinsic
excitations of the system.
The coefficients $D_1^{(0)}(q_1,q_1,q'_1,t)$ and
$D_2^{(0)}(q_1,q_1,q'_1,q'_2,t)$ being of very similar structure we,
henceforth, shall refer to $D_1^{(0)}(q_1,q_1,q'_1,t)$ only, as the results
obtained for $D_1^{(0)}$ can be immediately transferred to $D_2^{(0)}$.

The purpose of the following investigation is to show that a
fluctuation-dissipation relation holds for these parts of the coefficients
$D_1$ and $D_2$.

To start with, we remind the reader of the fact that the coefficients
$D_1^{(0)}$ and $D_2^{(0)}$ are defined by the parts of order $\varepsilon^0$
in the overlap parameter $\varepsilon$. Consequently, by consistency, we have
to put
\begin{align}
\langle aq\vert\delta \hat{H}^{I}_{cpl}\,
\delta \hat{H}_{cpl}\vert aq  \rangle =&\sum_{a'q'}
\langle aq\vert\delta \hat{H}^{I}_{cpl}\vert a'q'  \rangle
\langle a'q'\vert \delta \hat{H}_{cpl}\vert aq  \rangle  \notag\\
\simeq & \sum_{a'}
\langle aq\vert\delta \hat{H}^{I}_{cpl}\vert a'q  \rangle
\langle a'q\vert \delta \hat{H}_{cpl}\vert aq  \rangle
  \label{5.17}
\end{align}
Let us introduce the Fourier transform $\widetilde{\chi}(q,\omega;t)$ of the
response function $\chi(q,s';t)$
\begin{equation}\label{5.18}
\widetilde{\chi}(q,\omega;t) :=\int_{-\infty }^{+\infty } ds' \chi(q,s';t)
e^{i\omega s'} ,
\end{equation}
the so-called ``\underline{dynamical susceptibility}", and its
``\underline{dissipative}" part $\widetilde{\chi}^d(q,\omega;t)$ defined by
\begin{equation}\label{5.19}
\widetilde{\chi}^d(q,\omega;t)=\frac{1}{2i}
\left(\widetilde{\chi}(q,\omega;t) - \widetilde{\chi}(q,-\omega;t)\right)
\end{equation}
It is also given by
\begin{equation}\label{5.20}
\widetilde{\chi}^d(q,\omega;t)=\int_{-\infty }^{+\infty } ds' \chi(q,s';t)
\sin(\omega s')
\end{equation}
It is easily seen that $\widetilde{\chi}^d$ can be written in the form
\begin{align}
\widetilde{\chi}^d(q,\omega;t)=\frac{1}{2\hbar} \int_{-\infty }^{+\infty }
\!\!\! ds' \sum_{a} F_a(q,T(q,t))
\langle aq\vert \left[\delta \hat{H}^{I}_{cpl}(s',t),
\delta \hat{H}_{cpl}(t)\right]_-\vert aq \rangle  e^{i\omega s'}
  \label{5.21}
\end{align}
The important point is that no Heaviside functions appear any more in the
representation (\ref{5.21}) of the dissipative susceptibility. Due to this
feature, the r.h.s. of (\ref{5.21}) is proportional to the Fourier transform
of the antisymmetrical part $A(q,s';t)$ of the function
\begin{equation}\label{5.22}
C(q,s';t):= \sum_{a} F_a(q,T(q,t))
\langle aq\vert \delta \hat{H}^{I}_{cpl}(s',t)
\delta \hat{H}_{cpl}(t)\vert aq \rangle
\end{equation}
which represents the correlation between the fluctuations
$\delta \hat{H}_{cpl}(t) \!\!= \!\!\delta \hat{H}^{I}_{cpl}(0,t)$ and
$ \delta \hat{H}^{I}_{cpl}(s',t)$.

Analogously to the decomposition of the expressions (\ref{5.11}),
(\ref{5.12}), we write (\ref{5.22}) in the form
\begin{equation}\label{5.23}
C(q,s';t) = S(q,s';t) +i A(q,s';t),
\end{equation}
where the symmetrical and the antisymmetrical parts of the correlation
function $C$ are defined by
\begin{align}
&S(q,s';t) := \frac{1}{2} \sum_{a} F_a(q,T(q,t))
\langle aq\vert \left[\delta \hat{H}^{I}_{cpl}(s',t),
\delta \hat{H}_{cpl}(t)\right]_+\vert aq \rangle
  \label{5.24}\\
&A(q,s';t) :=\frac{1}{2i}\sum_{a} F_a(q,T(q,t))
\langle aq\vert \left[\delta \hat{H}^{I}_{cpl}(s',t),
\delta \hat{H}_{cpl}(t)\right]_-\vert aq \rangle
  \label{5.25}
\end{align}
The Fourier transforms, sometimes referred to as ``\underline{spectral
functions}", are given by
\begin{equation}\label{5.26}
\widetilde{C}(q,\omega;t)=\int_{-\infty }^{+\infty } ds' C(q,s';t)
e^{i\omega s'} =\widetilde{S}(q,\omega;t) + i\widetilde{A}(q,\omega;t),
\end{equation}
where $\widetilde{S}(q,\omega;t)$ and $\widetilde{A}(q,\omega;t)$ are the
Fourier transforms of $S(q,\omega;t)$ and  $A(q,\omega;t)$.
Comparing the r.h.s. of Eq. (\ref{5.21}) with the definition of
$\widetilde{A}(q,\omega;t)$, one obtains the relation
\begin{equation}\label{5.27}
\widetilde{\chi}^d(q,\omega;t)=\frac{i}{\hbar} \widetilde{A}(q,\omega;t)
\end{equation}
which is one way of writing the fluctuation-dissipation relation.

In what follows, we shall show that the r.h.s. of Eq. (\ref{5.27}) can be
written in the form
\begin{equation}\label{5.28}
\frac{i}{\hbar} \widetilde{A}(q,\omega;t) =\frac{1}{2\hbar}
\left(1-e^{-\beta\hbar \omega}\right) \widetilde{C}(q,\omega;t)
\end{equation}
where $\beta(q,t)$ is the reciprocal temperature
\begin{equation}\label{5.29}
\beta(q,t):=\frac{1}{T(q,t)}
\end{equation}
The proof of the relation (\ref{5.28}) is based on the ``stationarity" of the
mean-value
\begin{align}
\langle \langle \delta \hat{H}^{I}_{cpl}(s',t)
\delta \hat{H}_{cpl}(t)\rangle\rangle_q
= \langle \langle \delta \hat{H}_{cpl}(t)
\delta \hat{H}^{I}_{cpl}(-s',t) \rangle\rangle_q
  \label{5.30}
\end{align}
and on the following symmetry relation for the correlation function
\begin{equation}\label{5.31}
C(q,s';t) = C(q,-s'\!\!-i\hbar \beta;t)
\end{equation}
Although these two relations are well-known (see for instance Ref.
\cite{Grossmann}), we have to discuss their validity, which is not automatic in
the context of our theory:  \\
As a prerequisite of the proof, we have to accept the approximation
(\ref{5.17}) of only considering terms of order $\varepsilon^0$. We can thus
use the cyclic invariance of the trace in a first step
\begin{align}
\langle \langle \delta \hat{H}^{I}_{cpl}(s',t)
\delta \hat{H}_{cpl}(t)\rangle\rangle_q
:=& \!\! \sum_{a} \langle aq\vert
\frac{e^{-\beta(q) \hat{H}'_1(q)}}{Z_0(q)}
e^{\frac{i}{\hbar}s'\hat{H}_1}    \delta \hat{H}_{cpl}(t)
e^{-\frac{i}{\hbar}s'\hat{H}_1}    \delta \hat{H}_{cpl}(t)
\vert aq  \rangle \notag\\
=&  \sum_{a}  \langle aq\vert
e^{-\frac{i}{\hbar}s'\hat{H}_1}    \delta \hat{H}_{cpl}(t)
\frac{e^{-\beta(q) \hat{H}'_1(q)}}{Z_0(q)}
e^{\frac{i}{\hbar}s'\hat{H}_1}    \delta \hat{H}_{cpl}(t)
\vert aq  \rangle
  \label{5.32}
\end{align}
ans, subsequently, the commutation property
\begin{equation}\label{5.33}
\left[\hat{H}_1,\hat{H}'_1(q)\right]=0
\end{equation}
in a second step, for rewriting the r.h.s. of Eq. (\ref{5.32})
\begin{align*}
& =\sum_{a} \langle aq\vert \delta \hat{H}^{I}_{cpl}(-s',t)
\frac{e^{-\beta(q) \hat{H}'_1(q)}}{Z_0(q)}
  \delta \hat{H}_{cpl}(t) \vert aq  \rangle
=  \langle \langle \delta \hat{H}_{cpl}(t)
\delta \hat{H}^{I}_{cpl}(-s',t) \rangle\rangle_q
\end{align*}
This proves relation (\ref{5.30}).\\
Let us now consider the explicit form of the function
$ C(q,-s'\!\!-i\hbar \beta;t)$
\begin{align}
& C(q,-s'\!\!-i\hbar \beta;t)= \sum_{a} \langle aq\vert
\frac{e^{-\beta(q) \hat{H}'_1(q)}}{Z_0(q)}
 e^{\beta(q) \hat{H}_1} \delta \hat{H}^{I}_{cpl}(-s',t)
e^{-\beta(q) \hat{H}_1}    \delta \hat{H}_{cpl}(t)
\vert aq  \rangle
  \label{5.34}
\end{align}
If the Hamiltonian $\hat{H}'_1(q)$, which appears in our equilibrium
distribution, were equal to the Hamiltonian $\hat{H}_1$ defining the
interaction picture, the r.h.s. of Eq. (\ref{5.34}) would be equal to
$C(q,s',t)$ because of the cyclic invariance.\\
In our case, these two Hamiltonians differ because we do not consider a true
thermal equilibrium but a constrained one.
In the expression (\ref{5.34}), we may replace the operator $\hat{H}_1$
(see Eq. (\ref{3.50'})) by $\hat{H}_1(q)$, i.e. by the part of
$\hat{H}_1$ acting on the subspace of states $\vert aq\rangle$ with fixed
value $q$ :
\begin{equation}\label{5.35}
\hat{H}_1(q) =\sum_{a} \vert aq\rangle \left( \rule{0mm}{3mm}
{\cal E}'_a(q)+\lambda(q)
\langle aq\vert \hat{q}\vert aq\rangle\right) \langle aq \vert.
\end{equation}
The Hamiltonian $\hat{H}'_1(q)$
\begin{equation}\label{5.36}
\hat{H}'_1(q) =\sum_{a} \vert aq\rangle {\cal E}'_a(q) \langle aq \vert
\end{equation}
differs from $\hat{H}_1(q)$ by $\lambda(q) \hat{q}(q)$
\begin{equation}\label{5.37}
\hat{H}_1(q)=\hat{H}'_1(q)  + \lambda(q) \hat{q}(q)
\end{equation}
$\hat{q}(q)$  being defined by
\begin{equation}\label{5.38}
\hat{q}(q):=\sum_{a'}\vert a' q\rangle \langle a' q\vert \hat{q}
\vert a'q\rangle \langle a'q \vert
\end{equation}
Eq. (\ref{5.34}) can thus be written in the form
\begin{align}
C(q,-s'\!\!-i\hbar \beta;t)=& \sum_{a} \langle aq\vert
\frac{1}{Z_0(q)} e^{\beta(q)\lambda(q) \hat{q}(q)}
 \delta \hat{H}^{I}_{cpl}(-s',t)
e^{-\beta(q)\left(\hat{H}'_1(q)  + \lambda(q) \hat{q}(q)\right)}
 \delta \hat{H}_{cpl}(t) \vert aq  \rangle \notag\\
=& \sum_{a} \langle aq\vert \delta \hat{H}^{I}_{cpl}(-s',t)
\frac{e^{-\beta(q) \hat{H}'_1(q)}}{Z_0(q)}
e^{-\beta(q)\lambda(q) \hat{q}(q)}   \delta \hat{H}_{cpl}(t)
e^{\beta(q)\lambda(q) \hat{q}(q)} \vert aq  \rangle
  \label{5.39}
\end{align}
In Eq. (\ref{5.39}), the matrix-elements
\begin{align*}
&\langle a'q\vert
e^{-\beta(q)\lambda(q) \hat{q}(q)}   \delta \hat{H}_{cpl}(t)
e^{\beta(q)\lambda(q) \hat{q}(q)} \vert aq  \rangle =
\langle a'q\vert \hat{H}_{cpl}(t)  \vert aq  \rangle
e^{-\beta(q)\lambda(q)
\left(\langle a'q\vert \hat{q}\vert a'q \rangle -
\langle aq\vert \hat{q}\vert aq \rangle\right)}
\end{align*}
appear. The symmetry (\ref{5.31}) of the correlation function $C$ then holds,
if we may put the exponential factor approximately equal to 1
\begin{equation}\label{5.40}
e^{-\beta(q)\lambda(q)
\left(\langle a'q\vert \hat{q}\vert a'q \rangle -
\langle aq\vert \hat{q}\vert aq \rangle\right)} \simeq 1
\end{equation}
The approximation (\ref{5.40}) holds, if the multipole moments in different
intrinsic excitations do not fluctuate too much.

Let us assume that the approximation (\ref{5.40}) is valid. We then claim
that the following relation holds for the Fourier transforms of the
correlation function
\begin{equation}\label{5.41}
\widetilde{C}(q,\omega;t)=\widetilde{C}(q,-\omega;t)e^{\beta(q)\hbar \omega}
\end{equation}
as a result of the symmetry property (\ref{5.31}). The proof is
straightforward:
\begin{align*}
\widetilde{C}(q,\omega;t)=&
\int_{-\infty }^{+\infty } ds' C(q,s';t) e^{i\omega s'}
\stackrel{(\ref{5.31})}{=}
\int_{-\infty }^{+\infty } ds' C(q,-s'\!\!-i\hbar \beta;t) e^{i\omega s'}\\
=&\int_{-\infty-i\hbar \beta}^{+\infty-i\hbar \beta} ds''
C(q,s'';t) e^{-i\omega s''}e^{\beta(q)\hbar \omega}
= \widetilde{C}(q,-\omega;t)e^{\beta(q)\hbar \omega}
\end{align*}
As an immediate consequence of the relation (\ref{5.41}) we obtain
\begin{align}
&\widetilde{S}(q,\omega;t)=\frac{1}{2}
\left(1+ e^{-\beta(q)\hbar \omega} \right)
\widetilde{C}(q,-\omega;t)
  \label{5.42}\\
&\widetilde{A}(q,\omega;t)=\frac{1}{2i}
\left(1- e^{-\beta(q)\hbar \omega} \right)
\widetilde{C}(q,-\omega;t)
  \label{5.43}
\end{align}
Substituting the result (\ref{5.43}) into Eq. (\ref{5.27}), we find the
conventional form
\begin{equation}\label{5.44}
\widetilde{\chi}^d(q,\omega;t)=\frac{1}{2\hbar }
\left(1- e^{-\beta(q)\hbar \omega} \right)
\widetilde{C}(q,-\omega;t)
\end{equation}
of the fluctuation-dissipation relation \cite{Grossmann}.

In Section \ref{introduction}, we had mentioned the simple classical result
of Einstein (see Eq. (\ref{1.1})) relating the diffusion coefficient $D$ to
the friction constant $\gamma$ for the case of Brownian particles of mass $M$
immersed into a gaseous or liquid medium.\\
The classical limit of the relation (\ref{5.44}) is obtained in the limit
\begin{equation}\label{5.45}
\beta(q,t)\hbar \omega=\frac{\hbar \omega}{T(q,t)} \ll 1
\end{equation}
In the approximation
\begin{equation}\label{5.46}
e^{-\beta\hbar \omega}\simeq 1-\beta\hbar \omega
\end{equation}
the fluctuation-dissipation relation (\ref{5.44}) assumes the form
\begin{equation}\label{5.47}
\widetilde{C}(q,-\omega;t)
\simeq \frac{2T}{ \omega } \widetilde{\chi}^d(q,\omega;t)
\end{equation}
It does no longer contain explicitly Planck's constant and corresponds to the
Einstein relation (\ref{1.1}).

Using Eq. (\ref{5.42}), one can also relate the Fourier transform
$\widetilde{S}(q,\omega;t)$ to $\widetilde{\chi}^d(q,\omega;t)$:
\begin{align}
& \widetilde{S}(q,\omega;t)=\hbar \coth
\left(\frac{\beta(q,t)\hbar \omega}{2}\right)
\widetilde{\chi}^d(q,\omega;t)
  \label{5.48}
\end{align}
with the classical limit
\begin{align}
& \widetilde{S}(q,\omega;t)\simeq \frac{2T}{ \omega }
\widetilde{\chi}^d(q,\omega;t)
  \label{5.49}
\end{align}
It is thus shown that the parts of order O($\varepsilon^0$) of the
coefficients $D_1$ and $D_2$ (eqs. (\ref{5.2}), (\ref{5.3})) do fulfill a
fluctuation-dissipation relation.
This is expected from the fact that in the limit O($\varepsilon^0$), the
coefficients $D_1$ and $D_2$ do not describe a transport of the system
between different values of the generator coordinates but rather a
quasi-stationary equilibrium at a given value of the generator coordinates.

\section{Simple limits of the theory}          \label{limits}  
\setcounter{equation}{0}

Let us consider the equation of motion (\ref{3.77}) for the reduced density
matrix $R(q_1,q_2;t)$ in some special cases:

If we restrict the Hill-Wheeler space to the grounstates $\phi_{a_0}(q)$ at each
given value of the g.c. $q$, eq. (\ref{3.77}) assumes the simple form
\begin{align}
&\frac{dR(q_1, q_2 ; t)}{dt}=
-\frac{i}{\hbar} \sum_{q_3}
 \left[\rule{0mm}{3mm}{\cal H}_{a_0a_0}(q_1,q_3) R(q_3,q_2;t)
 - R(q_1,q_3;t)  {\cal H}_{a_0a_0}(q_3,q_2) \right]
\label{6.1}
\end{align}
This equation is equivalent to the description of the fission process by the
usual GCM without intrinsic excitations. The latter theory underlies the
calculations of Refs. \cite{Berger-Gogny,Goutte}. Using this theory, it was
found that, at small excitation energies ($\lesssim 10$ MeV), the calculated
mass distribution of the fragments in $^{238}$U agreed quite well with
the experimental one, although its width was a little too small \cite{Goutte}.
In this respect, it should be noted that the omission of intrinsic excitations
does not mean that vibrations of the nucleus shape on the way to scission are
neglected. On the contrary, the dynamical role of these vibrations was proved to
be important to explain the large width of the calculated mass distribution.

Another approximate version of the equation of motion (\ref{3.77}) is obtained
in the case that the ``local'' temperature $T(q,t)$ is small compared to the
energy of the lowest elementary excitation $a\neq a_0$:
\begin{equation}\label{6.2}
T(q,t) \ll {\rm Min}\left(\rule{0mm}{3mm}{\cal E}'_a(q)
-{\cal E}'_{a_0}(q)\right)
\end{equation}
The Boltzmann factors $F_a(q,T)$ can be written in the form
\begin{align}
F_a(q,T)=&\frac{e^{-{\cal E}'_a(q)/T}}{Z_0(q,T)}
=\frac{e^{-\left({\cal E}'_a(q)-{\cal E}'_{a_0}(q)\right)/T}}
{1+\sum_{b\neq a_0}e^{-\left({\cal E}'_b(q)-{\cal E}'_{a_0}(q)\right)/T}}
  \label{6.3}
\end{align}
which, in the limit of Eq. (\ref{6.2}), yields
\begin{equation}\label{6.4}
F_a(q,T)\simeq \delta_{aa_0}
\end{equation}
Using the same technique, it is easy to verify that the time-derivative
$\dot{F}_a=\dot{T}\,\partial F/\partial T$ goes to zero in the same limit.
If we substitute these limiting values into (\ref{3.77}), we obtain
\begin{align}
\frac{dR(q_1, q_2 ; t)}{dt}=&
-\frac{i}{\hbar} \sum_{q_3}
 \left[{\cal H}_{a_0a_0}(q_1,q_3) R(q_3,q_2;t)
 - R(q_1,q_3;t) {\cal H}_{a_0a_0}(q_3,q_2)\rule{0mm}{3mm} \right]
\notag\\&\hspace{-12mm}
-\frac{1}{\hbar^2}  \int_0^{\infty} ds \, \theta_0(t-s) \sum_{q_3q_4}
\sum_{a\neq a_0} \left\{
{{{\cal H}_{cpl}}_{a_0a}(q_1q_3;t) {\cal H}^{I}_{cpl}}_{aa_0}(q_3q_4;s-t,t)
e^{i\omega_{a_0a_0}(q_4,q_2)(s-t)} R(q_4,q_2;t)
\rule{0mm}{4mm}\right. \notag\\ & \left.\rule{0mm}{4mm} \hspace{25mm}
- {{\cal H}_{cpl}}_{aa_0}(q_1q_3;t) {{\cal H}^{I}_{cpl}}_{a_0a}(q_4q_2;s-t,t)
 e^{i\omega_{a_0a_0}(q_3,q_4)(s-t)} R(q_3,q_4;t)   \right\}
\label{6.5}
\end{align}
It is interesting to observe that, in the limit (\ref{6.2}), our equation of
motion (\ref{6.5}) still contains terms describing ``virtual'' transitions to
intrinsically excited states $\vert a q\rangle\neq \vert a_0 q\rangle$.
Since the two terms involving these transitions are of similar structure but of
different sign, we do not expect a large contribution from them. This is no
longer expected whenever the temperature $T(q,t)$ is no longer small compared to
the elementary excitations. In this case, the equation of motion (\ref{3.77})
must be solved as it stands, together with the eq. (\ref{4.24}) which determines
the temperature $T(q,t)$.

One notices from Eq. (\ref{6.3}) that the Boltzmann factor $F_a(q,t)$ becomes
very small whenever the intrinsic excitation energy
${\cal E}'_a(q)-{\cal E}'_{a_0}(q)$ is much larger than the temperature
$T(q,t)$. This means that, in the equation of motion (\ref{3.77}), only those
intrinsic excitations $\vert aq \rangle$ need to be taken into account which
correspond to non-negligible Boltzmann factors $F_a(q,t)$.

The actual computing work concerns the calculation of the matrix-elements
${\cal H}_{aa}(q,q')$ and ${{\cal H}_{cpl}}_{aa'}(q,q')$, on the one hand, and
the numerical solution of the integro-differential equation (\ref{3.77}), on the
other. One could convert the integral part of the equation into
partial differential terms of the Fokker-Planck type by a moment expansion of
the integral kernels. We do not think that this is useful, because the moment
expansion introduces approximations and the resulting partial differential
equation for $R(q_1,q_2;t)$ is also difficult to solve numerically. Thus we
think that a straightforward numerical solution of Eq. (\ref{3.77}) on the basis
of a discretization of the g.c. $q$ is the best way to carry out practical
applications of the theory.

Let us add that the techniques used in earlier calculations where intrinsic
excitations were omitted \cite{Berger-Gogny, Goutte} could be employed also in
actual applications of the present theory. For instance, in fission problems, it
will probably be convenient to define a ``scission region'' beyond which
fission fragments no longer exchange nucleons or internal energy and are
subject only to Coulomb repulsion and decay. Then fragments characteristics such as
mass--, charge-- or kinetic energy distributions (prior to neutron and
$\gamma$--ray emission) could be evaluated from the values
taken at large times by the solution of the equation of motion (\ref{3.77})
obtained in the scission region or slightly beyond.

\section{Summary and Discussion}          \label{summary}  
\setcounter{equation}{0}

{\bf a) \underline{Short review of what has been done}}

The basis of our work is the method proposed by Hill and Wheeler
\cite{Hill-Wheeler} of expanding the nuclear many-body state in terms of a
set of basis functions which depend parametrically on deformation variables
$q$ (see Eq. (\ref{1.4})) which are integrated over. In this way, the
description of the nuclear shape and of its dynamical change as a function of
time does not necessitate the introduction of superfluous degrees of
freedom.

This method which has been successfully applied to the description of single
nuclear states has been applied to the nuclear density operator and its time
evolution in the present paper.

We used as basis functions states of independently moving quasi-particles
defined as eigenstates $\phi_a(q)$ of the one-body part of
$\hat{H}'_1(q)$ of the constrained Hartree-Fock-Bogoliubov (HFB)
Hamiltonian (see Eq. (\ref{1.10})). It is a most welcome property of these
states that, in the limit of vanishing collective large scale motion, they
represent a self-consistently determined state of equilibrium at each given
shape of the nucleus.

A complication is produced by the non-orthogonality of these states. We coped
with it by either using a biorthogonal set of basis functions (see Eq.
(\ref{1.15}) and Section \ref{choice}) or by introducing an orthogonalized
version of the Hill-Wheeler states (see Eqs. (\ref{2.11}), (\ref{2.12})).

The basic assumptions of our theory are the ansatz (\ref{3.7}) for the slowly
time-dependent part $\hat{\rho}_P(t)$ of the statistical operator and the
Markov approximation (see text after Eq. (\ref{3.64})) simplifying the
equation of motion (\ref{3.64}) for $\hat{\rho}_P(t)$.
The physical idea underlying the choice of $\hat{\rho}_P(t)$ is that this
slowly time-dependent part of the statistical operator depends neither on the
detailed phase relations of the rapidly time-depending part of
$\hat{\rho}(t)$ nor on the exact occupation amplitudes of the intrinsic
excitations at a given shape of the nucleus which thus may be represented by
a canonical distribution (see Eq. (\ref{3.8})).

In Section \ref{determination}, the temperature $T$ which occurs in the
canonical distribution is chosen by the requirement that the reference
density $\hat{\rho}_P(t)$ should correctly describe the total mean energy
(see Eq. (\ref{4.1})) and the energy at a given shape $q$ of the nucleus (see
Eq. (\ref{4.3})) at all times.

An important question is whether our theory implies a relation between the
quantities describing the dissipation and fluctuation in the equation of
motion (\ref{3.76}) for $\hat{\rho}_P(t)$. We dealt with this question in
Section \ref{fluctuation}.

{\bf b) \underline{Short discussion of what should still be done}}

Obviously, a pertinent question is how one is to describe the emission of
nucleons and photons during the fission process. We note that the emission
during the dynamical process of fission is more difficult to treat than the
emission from the fragments which have reached a thermal equilibrium. We are
working on a description of emission processes during fission in the
framework of our theory. The work will be submited for publication at our
earliest convenience.

A very important endeavour will be to perform actual numerical calculations
on
the basis of the presented theory. Indeed, it is one of the purposes of the
microscopic theory to find out whether we are going to find a satisfactory
agreement between calculated and measured results on the fission process.
This is by no means obvious because not only the assumptions of our theory
are at stake but also the very concept of whether it is at all possible to
describe all  the low energy phenomena of nuclear physics on the basis of one
given effective nucleon-nucleon interaction. This implication is a much
farther going assumption than the hypothesis of the existence of an
energy-density functional enabling us to calculate ground state properties.

It should be noted that calculations on the fission process at very low
excitation energy have already been performed within the generator coordinate
method without inclusion of intrinsic excitations \cite{Goutte} where an
encouragingly good agreement between calculated and measured results has been
found. The present theory would give the possibility to extend these
calculations to higher excitation energy and compare them with the existing
careful experimental work.

Finally, a problem of its own merit would be to investigate in detail the
classical limit of the theory presented. It surely is related to the
$\delta$-function limit of the narrow gaussian overlap of many-body
wavefunctions pertaining to different values of the generator coordinates.
This fact suggests that the Fokker-Planck approximation of our equation of
motion (\ref{3.77}) for the reduced density $R(q_1,q_2;t)$ might be the
appropriate starting point for this investigation.

\vskip 5mm
{\bf Acknowledgements}

The collaboration between the authors of this paper was made possible by the
enduring hospitality extended to one of us (K.D.) by the Centre DAM \^Ile de
France de Bruy\`eres-le-Ch\^atel. K.D. expresses his gratitude to the Centre.

\vskip 5mm
{\bf Appendices}

\appendix

\section{Some properties of the projection operators} \label{appendix1}
\setcounter{equation}{0}

In Eq. (\ref{3.16}), we have introduced the matrix-representation of the
projection $\hat{P}$. As the temperature $T$ depends on time, the
operator $\dot{\hat{P}}$ is unequal from 0 and its matrix-representation
is
\begin{equation}\label{a1}
\dot{\hat{P}}_{a_1q_1a_2q_2a_3q_3a_4q_4} =
\delta_{a_1a_2} \delta_{q_1q_3} \delta_{q_2q_4} \delta_{a_3a_4}
\dot{F}_{a_2}(q_2,T(q_2,t))
\end{equation}
The matrix-representation of $\hat{P} \dot{\hat{P}}$ is thus obtained
as
\begin{align}
&\sum_{a_3q_3a_4q_4}  \hat{P}_{a_1q_1a_2q_2a_3q_3a_4q_4}
\dot{\hat{P}}_{a_3q_3a_4q_4a_5q_5a_6q_6} =\notag\\
&\hspace{10mm}
\delta_{a_1a_2} \delta_{q_1q_5} \delta_{q_2q_6} \delta_{a_5a_6}
F_{a_2}(q_2,T(q_2,t))\sum_{b}\dot{F}_{b}(q_2,T(q_2,t))=0
  \label{a2}
\end{align}
where the r.h.s. vanishes because we have
\begin{align*}
  \sum_{b}F_{b}(q_2,T(q_2,t))=1
\end{align*}
On the other hand, for the matrix-representation of
$\dot{\hat{P}} \hat{P}$, one finds
\begin{align}
&\sum_{a_3q_3a_4q_4} \dot{\hat{P}}_{a_1q_1a_2q_2a_3q_3a_4q_4}
\hat{P}_{a_3q_3a_4q_4a_5q_5a_6q_6} =\notag\\
&\hspace{10mm}
=\delta_{a_1a_2} \delta_{q_1q_5} \delta_{q_2q_6} \delta_{a_5a_6}
\dot{F}_{a_2}(q_2,T(q_2,t))\sum_{b}F_{b}(q_2,T(q_2,t))\notag\\
&\hspace{10mm}
= \dot{\hat{P}}_{a_1q_1a_2q_2a_5q_5a_6q_6}
  \label{a3}
\end{align}
The results (\ref{a2}), and (\ref{a3}) can be written in the form
\begin{align}
&\hat{P} (t) \dot{\hat{P}}(t)=0
  \label{a4}\\
&\dot{\hat{P}}(t) \hat{P}(t)=\dot{\hat{P}}(t)
  \label{a5}
\end{align}
respectively.
As we have
\begin{equation}\label{a6}
\dot{\hat{Q}}=-\dot{\hat{P}}
\end{equation}
as a consequence of Eq. (\ref{3.17}), we find
\begin{align}
&\hat{P}(t) \dot{\hat{Q}}(t)=0=\left(1-\hat{Q}(t)\right)
 \dot{\hat{Q}}(t)
  \label{a7}\\
& \dot{\hat{Q}}(t) \hat{P}(t)= \dot{\hat{Q}}(t)
=\dot{\hat{Q}}(t) \left(1-\hat{Q}(t)\right)
  \label{a8}
\end{align}
from (\ref{a4}), and (\ref{a5}). Eqs. (\ref{a7}) and (\ref{a8}) imply the
relations
\begin{align}
&\hat{Q}(t) \dot{\hat{Q}}(t)= \dot{\hat{Q}}(t)
  \label{a9}\\
& \dot{\hat{Q}}(t) \hat{Q}(t)= 0
  \label{a10}\\
&\hat{Q}(t) \dot{\hat{P}}(t)= \dot{\hat{P}}(t)
  \label{a11}\\
& \dot{\hat{P}}(t) \hat{Q}(t)= 0
  \label{a12}
\end{align}
For any given operator $\hat{A}$, the matrix-representation of
$\dot{\hat{P}}\hat{A}$ reads
\begin{align}
\langle a_1q_1\vert \dot{\hat{P}}(t) \hat{A}\vert a_2q_2 \rangle
&= \sum_{a_3q_3a_4q_4} \dot{\hat{P}}_{a_1q_1a_2q_2a_3q_3a_4q_4}
\langle a_3q_3\vert \hat{A}\vert a_4q_4 \rangle \notag\\
&\hspace*{-2mm}\stackrel{\rm (\ref{a1})}{=} \delta_{a_1a_2}
\dot{F}_{a_2}(q_2,T(q_2,t))\sum_{b}
\langle b q_1\vert \hat{A}\vert b q_2 \rangle \notag\\
&= \delta_{a_1a_2} A^{red}(q_1,q_2)
\frac{\partial F_{a_2}(q_2,T(q_2,t))}{\partial T}
\dot{T}(q_2,t)
  \label{a13}
\end{align}
where the ``reduced part" $A^{red}(q_1,q_2)$ of the matrix
$\langle a_1q_1\vert \hat{A}\vert a_2q_2 \rangle$ is defined by
\begin{equation}\label{a14}
A^{red}(q_1,q_2):= \sum_{b_1b_2}
\langle b_1q_1\vert \hat{A}\vert b_2q_2 \rangle \delta_{b_1b_2}
\end{equation}
The operator $\dot{\hat{P}}(t) \hat{A}$ is thus seen to be diagonal
with respect to the quantum numbers of intrinsic excitations and slowly
time-dependent as the temperature $T(q_2,t)$ depends slowly on time. The
factor $\partial F_{a_2}(q_2,T(q_2,t))/\partial T$ can be reformulated as in
Eq. (\ref{3.46}) and differs from a canonical distribution.

\section{Detailed derivation of the Green function (\ref{3.40})}
\label{appendix2}
\setcounter{equation}{0}
In what follows we show in detail how one obtains the solution (\ref{3.40})
of the equation (\ref{3.32}).
\begin{equation}\tag{3.32}
\frac{\partial \hat{G}(t,s)}{\partial t}+\frac{i}{\hbar}
\hat{Q}(t)\hat{L}\hat{G}(t,s)
=\hat{Q}(t)\delta(t-s)
\end{equation}
This equation implies that the Green operator $\hat{G}(t,s)$ can be written
in the form
\begin{equation}\label{b1}
\hat{G}(t,s)=\hat{Q}(t)\hat{K}(t,s)
\end{equation}
Sunstituting (\ref{b1}) into Eq. (\ref{3.32})
\begin{align}
\dot{\hat{Q}}(t)\hat{K}(t,s)+\hat{Q}(t)\frac{\partial\hat{K}(t,s)}
{\partial t} +\frac{i}{\hbar} \hat{Q}(t)\hat{L}\hat{Q}(t)\hat{K}(t,s)
=\hat{Q}(t)\delta(t-s)
  \label{b2}
\end{align}
and using the relations (see Appendix \ref{appendix1})
\begin{align*}
\dot{\hat{Q}}(t)=-\dot{\hat{P}}(t)=-\hat{Q}(t)\dot{\hat{P}}(t),
\end{align*}
we find that the Eq. (\ref{b2}) can be written in the form
\begin{align}
\hat{Q}(t)\left[ \frac{\partial\hat{K}(t,s)}{\partial t}
- \dot{\hat{P}}(t)\hat{K}(t,s)+\frac{i}{\hbar}\hat{L}\hat{Q}(t)
\hat{K}(t,s) -\delta(t-s)\right]=0
  \label{b3}
\end{align}
The operator $\hat{K}(t,s)$ is thus seen to satisfy the equation
\begin{equation}\label{b4}
\frac{\partial\hat{K}(t,s)}{\partial t} -\hat{W}(t)\hat{K}(t,s)=\delta(t-s)
\end{equation}
with
\begin{equation}\label{b5}
\hat{W}(t):= \dot{\hat{P}}(t) -\frac{i}{\hbar}\hat{L}\hat{Q}(t)
\end{equation}
For solving (\ref{b4}) in the range $t\geq s$, we put
\begin{equation}\label{b6}
\hat{K}(t,s)=\theta_0(t-s)\hat{\bar{K}}(t,s)
\end{equation}
with the initial condition
\begin{equation}\label{b7}
\hat{\bar{K}}(t,t)=1
\end{equation}
Substituting (\ref{b6}) into (\ref{b4}) yields
\begin{align}
\delta(t-s)\hat{\bar{K}}(t,s)+\theta_0(t-s)\left\{
\frac{\partial\hat{\bar{K}}(t,s)}{\partial t} - \hat{W}(t)
\hat{\bar{K}}(t,s)\right\}= \delta(t-s)
  \label{b8}
\end{align}
and, because of
\begin{equation}\label{b9}
\delta(t-s)\hat{\bar{K}}(t,s)=\delta(t-s)\hat{\bar{K}}(t,t)
\stackrel{(\ref{b7})}{=} \delta(t-s),
\end{equation}
we find the homogeneous differential equation
\begin{equation}\label{b10}
\frac{\partial\hat{\bar{K}}(t,s)}{\partial t} =
 \hat{W}(t) \hat{\bar{K}}(t,s)
\end{equation}
The equation (\ref{b10}) and the initial condition (\ref{b7}) show that
$\hat{\bar{K}}(t,s)$ can be written in the form
\begin{equation}\label{b11}
\hat{\bar{K}}(t,s) =1 +\int_{s}^{t} d\tau\, \hat{W}(\tau)
\hat{\bar{K}}(\tau,s)
\end{equation}
The integral equation (\ref{b11}) has the solution
\begin{equation}\label{b12}
\hat{\bar{K}}(t,s)=\hat{T}\left\{e^{\textstyle
\int_s^t d\tau \,\hat{W}(\tau)}\right\}
\end{equation}
where $\hat{T}$ is the time-ordering operator. The form (\ref{b12}) is thus
seen to be a short way to formulate the iterative solution of Eq.
(\ref{b11}). Finally, from (\ref{b1}), (\ref{b6}), and (\ref{b12}) we find
the form
\begin{equation}\label{b13}
\hat{G}(t,s)=\theta_0(t-s)  \hat{Q}(t)  \hat{T}\left\{e^{\textstyle
\int_s^t d\tau \,\hat{W}(\tau)}\right\}
\end{equation}
of the Green operator.

\section{Comments on the non-hermitian nature of $\hat{\rho}_P(t)$}
\label{appendix3}
\setcounter{equation}{0}
As we have already pointed out in Section \ref{derivation}, our ansatz
(\ref{3.7}) for the approximate form of the density operator is not
hermitian. The reader might ask the justified question whether it would not
be preferable to derive and solve an equation of motion for the hermitian
ansatz
\begin{equation}\label{c1}
\hat{\rho}^h_P(t):=\frac{1}{2}\left(\hat{\rho}_P(t)
+\hat{\rho}^\dagger_P(t)\right)
\end{equation}
rather than solving the equation of motion (\ref{3.76}) for
$\hat{\rho}_P(t)$ and subsequently use its hermitian part for the evaluation
of physical quantities.

If we use the orthogonal basis functions $\Psi_a(q)$ (see Section
\ref{choice}), the matrix representation of $\hat{\rho}^h_P(t)$ has the form
\begin{align}
\langle \Psi_{a_1}(q_1)\vert \hat{\rho}^h_P(t) \vert  \Psi_{a_2}(q_2)\rangle
=\frac{\delta_{a_1a_2}}{2}\left(F_{a_1}(q_1,T(q_1,t))+F_{a_2}(q_2,T(q_2,t))
\rule{0mm}{3mm}\right)R(q_1,q_2;t)
  \label{c2}
\end{align}
It is seen that (\ref{c2}) is close to the matrix representation (\ref{3.7})
of $\hat{\rho}_P(t)$, if the Boltzmann factors
$F_{a_1}(q_1,T(q_1,t))$ and $F_{a_2}(q_2,T(q_2,t))$
do not differ very much one from another for pairs of values of the g.c.
where the reduced density matrix $R(q_1,q_2;t)$ is substantially  different
from zero.
We expect that the cases where (\ref{c2}) differs substantially from the
non-hermitian case
\begin{align}
\langle \Psi_{a_1}(q_1)\vert \hat{\rho}_P(t) \vert  \Psi_{a_2}(q_2)\rangle
=\delta_{a_1a_2}F_{a_2}(q_2,T(q_2,t))R(q_1,q_2;t)
  \label{c3}
\end{align}
are more important for small excitation energies (and temperatures), where
the canonical ansatz is anyhow less well-founded.





\end{document}